\documentclass[12pt]{iopart}

\usepackage{graphicx}
\usepackage{iopams}

\begin{document}
\title[Measurement of the electron's electric dipole moment using YbF]{Measurement of the electron's electric dipole moment using YbF molecules: methods and data analysis}

\author{D~M~Kara, I~J~Smallman, J~J~Hudson, B~E~Sauer, M~R~Tarbutt and E~A~Hinds}
\address{Centre for Cold Matter, Blackett Laboratory, Imperial College London, Prince Consort Road, London SW7 2AZ, United Kingdom.}

\begin{abstract}
We recently reported a new measurement of the electron's electric dipole moment using YbF molecules [Nature \textbf{473}, 493 (2011)]. Here, we give a more detailed description of the methods used to make this measurement, along with a fuller analysis of the data. We show how our methods isolate the electric dipole moment from imperfections in the experiment that might mimic it. We describe the systematic errors that we discovered, and the small corrections that we made to account for these. By making a set of additional measurements with greatly exaggerated experimental imperfections, we find upper bounds on possible uncorrected systematic errors which we use to determine the systematic uncertainty in the measurement. We also calculate the size of some systematic effects that have been important in previous electric dipole moment measurements, such as the motional magnetic field effect and the geometric phase, and show them to be negligibly small in the present experiment. Our result is consistent with an electric dipole moment of zero, so we provide upper bounds to its size at various confidence levels. Finally, we review the prospects for future improvements in the precision of the experiment.

\end{abstract}

\maketitle

\section{Introduction}

An electron has a magnetic dipole moment associated with its spin. The extraordinarily precise measurement of this magnetic moment \cite{Hanneke(1)08} is a demanding test of quantum electrodynamics and a probe for any possible electron sub-structure. A measurement of the electron's {\it electric} dipole moment (EDM, $d_{\rm{e}}$) also tests the laws of physics. A non\,-\,zero permanent electric dipole moment of an electron violates time-reversal symmetry (T). To see that this is the case, consider how an electron changes under time-reversal: the spin direction is reversed whereas the electric dipole moment, a static property that measures how the electric charge is distributed, is unchanged. Since spin is the electron's only internal degree of freedom, either the EDM is zero or it is not zero and T-symmetry is violated. According to the CPT theorem, violation of T-symmetry is equivalent to violation of CP-symmetry, the combined symmetry of charge conjugation and parity inversion. The observation of CP-violation in the decays of neutral K- and B-mesons \cite{Christenson(1)64, BABAR, Belle} is incorporated into the standard model of particle physics via the complex phase that appears in the quark mixing matrix. As a result of this mixing, the Standard Model predicts a non-zero electron EDM, though the prediction is exceedingly tiny, $d_{\rm{e}} < 10^{-38}\,e\,\rm{cm}$ \cite{Pospelov(1)05}. CP-violation is also essential to explain the observed asymmetry between matter and antimatter in the universe \cite{Sakharov(1)67}, but the quark mixing of the Standard Model is unable to account for this asymmetry. Extensions of the Standard Model, most notably supersymmetric extensions, introduce new sources of CP-violation that could explain the observed matter-antimatter asymmetry. These new theories predict EDM values that are far greater than in the Standard Model, typically by some 10 orders of magnitude \cite{Bernreuther(1)91}, and within the sensitivity range of current and planned experiments \cite{Commins(1)10}.

Measurements of the electron EDM use heavy, paramagnetic atoms or molecules which effectively enhance the interaction of $d_{\rm{e}}$ with the applied electric field \cite{Sandars(1)65, Sandars(1)66}. For many years the most precise measurement was made using a beam of thallium atoms, culminating in the 2002 result which found $d_{\rm{e}}$ to be consistent with zero and set an upper bound of $|d_{\rm{e}}| < 16 \times 10^{-28} e\,\rm{cm}$ \cite{Regan(1)02}. It has long been known that, because of their much greater polarizability, polar molecules offer even higher sensitivity to $d_{\rm{e}}$ than atoms \cite{Sandars(1)67, Sandars(1)75}. However, some of this intrinsic advantage is offset by the relative difficulty of producing and detecting the required heavy, polar, paramagnetic molecules. Recently, we made a new measurement of the electron EDM using a beam of YbF molecules, and set a new upper bound \cite{Hudson(1)11}. Here we give a detailed account of this experiment, focussing on the method, the analysis of the data, and the evaluation of the systematic uncertainty.

\section{Method}

\subsection{Overview}\label{Sec:Overview}

\begin{figure}[tb]
\centering
\includegraphics[width=0.9\columnwidth]{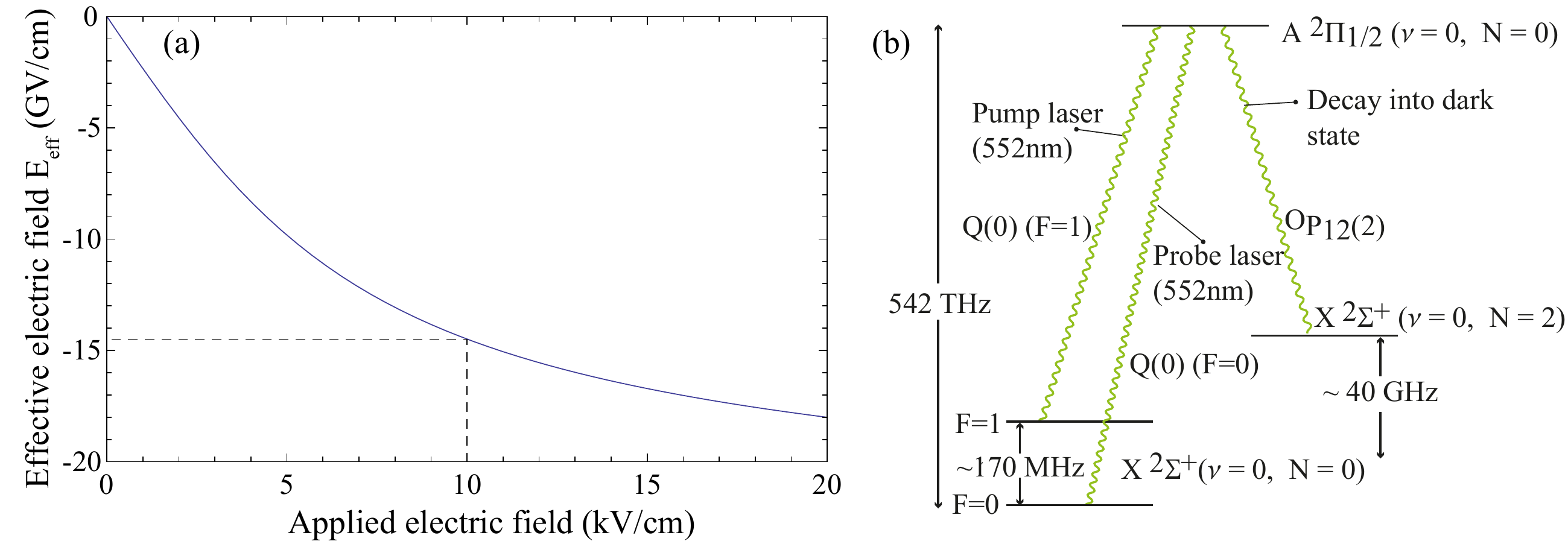}
\begin{center}
\caption{(a) The effective electric field for YbF as a function of the applied field. The dashed line shows the operating field in the experiment. (b) Relevant YbF energy levels and the transitions between them. \label{Fig:YbFInfo}}
\end{center}
\end{figure}

We measure the interaction energy between the EDM of the $^{174}\rm{YbF}$ molecule and an applied electric field $\vec{E}$, and interpret this as the interaction energy between the electron EDM, $\vec{d}$, and an effective electric field, $\vec{E}_{\rm{eff}}$. The EDM must lie along the symmetry axis defined by the spin, and so we write $\vec{d}=d_{\rm{e}}\,\vec{\sigma}$ where $\vec{\sigma}$ is a unit vector parallel to the spin. The effective electric field accounts for the polarization of the molecule in an applied field $\vec{E}$. This effective field is $\vec{E}_{\rm{eff}}=E_{\rm{eff}}^{\rm{max}} \eta(E) \hat{z}$ where $\hat{z}$ is a unit vector parallel to $\vec{E}$ and $\eta(E)=\langle\hat{n}\cdot\hat{z}\rangle$ is a polarization factor, $\hat{n}$ being a unit vector along the internuclear axis\footnote{Our convention is for $\hat{n}$ to point in the same direction as the molecular dipole moment, i.e. from the negative to the positive ion. Some papers in the field use a different convention, defining a unit vector that points from the heavy nucleus to the light one. For YbF, this is in the opposite direction to our $\hat{n}$.}, and the expectation value being evaluated using the eigenstate in the applied field. There are a number of calculations of $E_{\rm{eff}}^{\rm{max}}$ for YbF \cite{enhancement1, enhancement2, enhancement3, enhancement4, enhancement5, enhancement6}, with most results in agreement at the 10\% level. We take $E_{\rm{eff}}^{\rm{max}}=-26$\,GV\,cm$^{-1}$ \cite{enhancement2}. Figure \ref{Fig:YbFInfo}(a) plots $E_{\rm{eff}}$ versus $E$. The applied field in the experiment is $E=10$\,kV\,cm$^{-1}$, and for this field the polarization factor in the ground-state is $\eta=0.558$, giving $E_{\rm{eff}}=-14.5$\,GV\,cm$^{-1}$.

The electron EDM is not the only possible source of an interaction term proportional to $\vec{\sigma}\cdot \vec{E}$. There could also be P- and T-violating interactions between the electrons and nucleons, which would also give rise to a permanent EDM of the molecule, and are also sensitive to physics beyond the standard model \cite{Barr(1)92}. For YbF, the most important of these is a possible P,T-violating scalar-pseudoscalar electron-nucleon interaction \cite{Titov(1)06}. We have followed the usual convention of interpreting our result entirely in terms of an electron EDM.

The relevant energy levels of $^{174}$YbF are shown in Fig.\,\ref{Fig:YbFInfo}(b). We use only those molecules that are in the lowest rotational, vibrational and electronic state, $\rm{X}\,^2\Sigma^+\,(v=0,N=0)$. The interaction between the magnetic moments of the unpaired electron and the fluorine nucleus ($I=1/2$) splits the ground state into a pair of levels with total angular momentum quantum numbers $F=0$ and $F=1$, separated by approximately 170\,MHz. To measure the population in either hyperfine state we drive one of the two hyperfine components of the Q(0) transition to the short-lived electronically excited state $\rm{A}\,^2\Pi_{1/2}\,(v=0,N=0)$, and detect the resulting fluorescence. In an electric field $E \hat{z}$ the ground state of the molecule shifts down in energy. The shift is very nearly the same for both hyperfine components, but there are also some small differential shifts and these are particularly relevant in the experiment. The hyperfine splitting increases with increasing $E$ and the $|F,m_{F}\rangle=|1,0\rangle$ level splits away from the $|1,\pm 1\rangle$ levels. If the electron EDM is zero and there is no magnetic field applied, the $|1, \pm 1 \rangle$ levels are degenerate. A small magnetic field $B \hat{z}$ shifts these levels by $g \mu_{B} B m_{F}$, where the $g$-factor is very nearly 1, while a non-zero EDM results in the energy shift $-d_{e} E_{\rm{eff}} m_{F}$. Thus, a measurement of the electric-field induced splitting between these two $m_{F}$ levels measures the EDM. For the following discussion, we find it useful to define the states $|0\rangle=|0,0\rangle$, $|\pm 1\rangle=|1,\pm 1\rangle$, $|c\rangle = \frac{1}{\sqrt{2}}(|+1\rangle + |-1\rangle)$ and $|u\rangle = \frac{1}{\sqrt{2}}(|+1\rangle - |-1\rangle)$.

\begin{figure}[!tb]
\centering
\includegraphics[width=0.5\columnwidth]{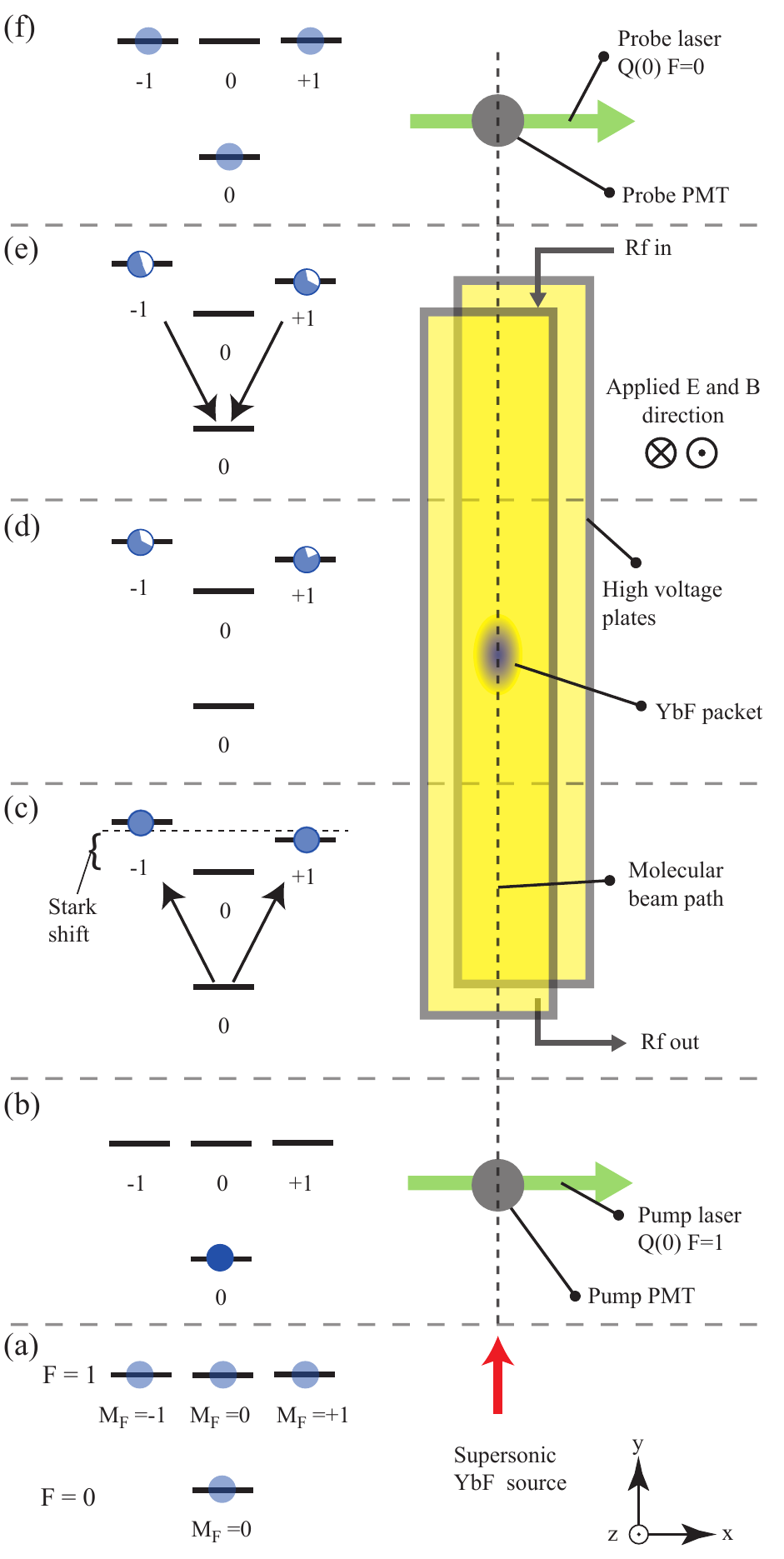}
\begin{center}
\caption{Overview of the experiment. (a) Pulses of YbF molecules emitted by the source with equal population in the 4 sublevels. (b) Population optically pumped out of $F=1$. (c) First rf pulse prepares an equal superposition of $M_{F}=+1$ and $M_{F}=-1$. (d) Phase difference accumulates due to applied $E$ and $B$ fields. (e) Second rf pulse transfers population back to $F=0$ with a probability that depends on the phase difference. (f) Population in $F=0$ probed by laser-induced fluorescence.}
\label{fig:overview}
\end{center}
\end{figure}

Figure \ref{fig:overview} gives an overview of the experiment. The molecular beam is inside a vacuum chamber and two layers of magnetic shielding. A detailed description of the apparatus is given in \cite{ApparatusPaper}. The source produces short pulses of cold YbF molecules that travel vertically upwards with a mean speed of 590\,m/s, taking about 2.2\,ms to traverse the length of the machine. We call each traversal of the machine by the molecules a `shot' of the experiment. The machine produces a shot every 40\,ms. In the following description of a shot we shall define the moment when the molecules are produced as the zero of time, $t=0$. The molecules first encounter the `pump' laser beam which propagates along $x$, is linearly polarized, and is tuned into resonance with the $F=1$ component of the Q(0) transition. This pumps the population out of $F=1$. About 40\% of this population is transferred to $F=0$. The rest is lost to the $\rm{X}\,^2\Sigma^+\,(v=0, N=2)$ state shown in Fig.\,\ref{Fig:YbFInfo}(b), or to higher-lying vibrational states in X, and no longer participates in the experiment. The fluorescence induced by the pump laser is detected on a photomultiplier tube (the `pump PMT') with 10\,$\mu$s time resolution, providing a measure of the molecule number in each shot.

The molecules next enter the interaction region defined by a pair of electric field plates, 75\,cm long, 7\,cm wide and 1.2\,cm apart, where static electric and magnetic fields $(E,B)\hat{z}$ are applied, typically with $E=\pm 10$\,kV/cm and $B=\pm 13.6$\,nT. The pair of plates also acts as a TEM transmission line for propagating 170\,MHz radiation in either direction. The geometry of this transmission line ensures that the rf magnetic field is linearly polarized along $\hat{x}$. It therefore couples the states $|0\rangle$ and $|c\rangle$, but does nothing to $|u\rangle$. At $t=1.1$\,ms, when the molecules are approximately 13\,cm inside the field plates, an 18\,$\mu$s-long rf pulse is applied with frequency tuned to the Stark-shifted $|0\rangle \leftrightarrow |c\rangle$ transition frequency, and amplitude optimized for driving a $\pi$-pulse, so that all the population is transferred from $|0\rangle$ to $|c\rangle$. During the subsequent free-evolution time of $T=642$\,$\mu$s, this state evolves into $\frac{1}{\sqrt{2}}(e^{i \phi}|1,+1\rangle + e^{-i \phi} |1,-1\rangle)=\cos\phi |c\rangle + i \sin\phi |u\rangle$ where

\begin{equation}\label{phi}
\phi = (g \mu_{B} B - d_{e} E_{\rm{eff}})T/\hbar.
\end{equation}
A second 18\,$\mu$s-long rf $\pi$-pulse is then applied, once again coupling $|c\rangle$ to $|0\rangle$, so that the final state is $\cos\phi |0\rangle + i \sin\phi |u\rangle$.

\begin{figure}[tb]
\centering
\includegraphics[width=0.5\columnwidth]{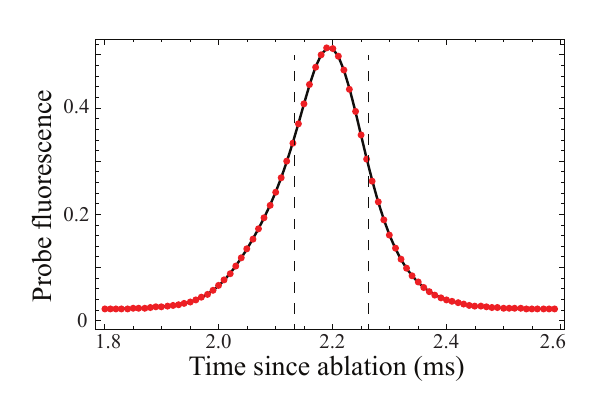}
\begin{center}
\caption{Fluorescence measured at the probe PMT as a function of arrival time. In analyzing EDM data we have used the part of the arrival-time distribution lying between the dashed lines.}
\label{fig:tof}
\end{center}
\end{figure}

Finally, the molecules pass through the linearly-polarized `probe' laser beam which propagates parallel to the pump beam and is tuned into resonance with the $F=0$ component of the Q(0) transition. The polarization directions of this beam and the pump beam are controlled by a pair of electronically rotatable polarizers. The resulting laser-induced fluorescence is detected on a second photomultiplier tube (the `probe PMT'), again with 10\,$\mu$s time resolution. This signal measures the final $|0\rangle$ population which is proportional to $\cos^{2}\phi$. Figure \ref{fig:tof} shows an example of the laser-induced fluorescence signal measured at the probe PMT as a function of time. The arrival-time distribution is approximately Gaussian, reflecting the velocity distribution of the molecules.

\subsection{Lineshape}\label{Sec:lineshape}

Now we derive a more general expression for how the signal depends on the experimental parameters. Our model contains only the states $|0\rangle$, $|+1\rangle$ and $|-1\rangle$, which (in this order) define a basis (let us call it the `z-basis'). The energies of these states are $-\hbar\Omega/2$, $\hbar (\Omega/2 + \Delta_{z})$ and  $\hbar (\Omega/2 - \Delta_{z})$ where $\Omega$ is the Stark-shifted hyperfine interval and $\hbar \Delta_{z} = g \mu_{B} B - d_{e} E_{\rm{eff}}$ is the sum of the Zeeman and EDM interaction energies for static magnetic and electric fields in the z-direction. The $|1,0 \rangle$ state plays no part in the experiment because it is Stark-shifted by approximately 8\,MHz from the $|\pm 1\rangle$ states, and because our $x$-polarized rf magnetic field cannot drive the transition from $|0\rangle$ to $|1,0 \rangle$. In the z-basis, the propagator that describes the free evolution of the state vector for a time $\tau$ is

\begin{equation}
\label{Eq:pi0} \Pi_{\rm{free}}(\tau)=\left(\begin{array}{lll}
e^{i\frac{\Omega}{2}\tau} & 0 & 0\\
0 & e^{-i(\frac{\Omega}{2}+\Delta_z)\tau} & 0\\
0 & 0 & e^{-i(\frac{\Omega}{2}-\Delta_z)\tau}\\
\end{array}\right).
\end{equation}

\noindent To find the propagator that describes how the state evolves in the rf magnetic field $B_{\rm{rf}}\cos(\omega t + \varphi)\hat{x}$, it is convenient to work in the `x-basis' defined by $|0\rangle$, $|c\rangle$ and $|u\rangle$. The two bases are related by the transformation

\begin{equation}\label{Eq:U}
U = U^{-1} = \left(\begin{array}{lll}
1 & 0 & 0\\
0 & \frac{1}{\sqrt{2}} & \frac{1}{\sqrt{2}} \\
0 & \frac{1}{\sqrt{2}}  & -\frac{1}{\sqrt{2}} \\
\end{array}\right).
\end{equation}

\noindent For a pulse of constant amplitude that turns on at time $t_{1}$ and off at $t_{1}+\tau$, and is detuned from resonance by $\delta=\omega-\Omega$, the solution of the time\,-\,dependent Schr\"{o}dinger equation gives the following propagator \cite{Ramsey(1)56}, written in the x-basis:

\begin{equation}\label{Eq:rfPropagator}
\Pi_{\rm{rf}}(t_1,\tau)=
\left(\begin{array}{lll}
Z e^{i\frac{\omega}{2}\tau} & W e^{i\frac{\omega}{2}\tau}e^{i(\omega t_1 + \varphi)} & 0\\
W e^{-i\frac{\omega}{2}\tau}e^{-i(\omega t_1 + \varphi)} & Z^* e^{-i\frac{\omega}{2}\tau} & 0\\
0 & 0 & e^{-i\frac{\Omega}{2} \tau}\\
\end{array}\right),
\end{equation}
where
\begin{equation}\label{Eq:ab}
\fl Z=\cos(\frac{a\tau}{2})-i\frac{\delta}{a}\sin(\frac{a\tau}{2}), \,\, W=-i\frac{b}{a}\sin(\frac{a\tau}{2}), \,\,a=\sqrt{\delta^2+b^2},\,\, b=\left<0\right|-\frac{\mu_x B_{\rm{rf}}}{\hbar}\left|c\right>\,.
\end{equation}

\noindent Here, $\mu_{x}$ is the $x$-component of the magnetic dipole moment operator. In order to obtain this propagator, we have applied the rotating-wave approximation and have neglected the Zeeman and EDM interactions by setting $\Delta_{z}=0$. In the experiment, the rf transitions are driven in the presence of the small applied magnetic field which rotates $|c\rangle$ into $|u\rangle$ at the angular frequency $\Delta_{z}$ during the excitation pulse. A numerical calculation that includes this \cite{Kara(1)10} shows that the effect on the lineshape is minor - as expected a slightly larger phase is acquired for a given $B_z$ because there is some phase evolution during the rf pulses, and the interference contrast is slightly reduced because of the Zeeman splitting.

The experiment is an interferometer consisting of two rf pulses with Rabi frequencies, durations and detunings $b_{1,2}$, $\tau_{1,2}$ and $\delta_{1,2}$, having a relative phase $\Phi_{\rm{rf}}$, and separated by a free evolution time $T$. The initial and final states, $|i\rangle$ and $|f\rangle$, both expressed in the z-basis, are related by the propagator
\begin{equation}\label{interferometerMatrix}
U \cdot \Pi_{\rm{rf2}}(t+\tau_{1}+T,\tau_{2}) \cdot U \cdot \Pi_{\rm{free}}(T) \cdot U \cdot \Pi_{\rm{rf1}}(t,\tau_{1}) \cdot U.
\end{equation}
Given that the pump laser prepares molecules in the initial state $|i\rangle=|0\rangle$, and the probe laser measures the final population in state $|0\rangle$, we find the signal to be
\begin{equation}\label{interferometerOutput}
S = |\langle 0|f \rangle|^{2} = S_{C} + S_{I} + S_{R}
\end{equation}
where
\begin{eqnarray}\label{Eq:Signals}
\fl S_{C}=\left({c_1}^{2}+\Delta_1^2{s_1}^{2}\right)\left({c_2}^{2}+\Delta_1^2{s_2}^{2}\right),\\
\fl S_{I}=\beta_{1}^{2} \beta_{2}^{2} s_{1}^{2} s_{2}^2\cos^{2}(\phi),\\
\fl S_{R}=2\beta_{1}\beta_{2}{s_1} {s_2}\cos(\phi)\left[
\left(
\Delta_{1}\Delta_{2} s_{1} s_{2}-c_{1} c_{2}
\right) \cos(\vartheta)
 +\left(
\Delta_{1} s_{1} c_{2} + \Delta_{2} c_{1} s_{2}
\right)\sin(\vartheta)\right],
\end{eqnarray}
and we have used the shorthand notation
\begin{equation}
c_i=\cos\left(\frac{a_i\tau_i}{2}\right), \, s_i=\sin\left(\frac{a_i\tau_i}{2}\right), \, \Delta_{i}=\delta_{i}/a_{i}, \,\,\, \beta_{i}=b_{i}/a_{i}.\nonumber
\end{equation}

\noindent $S_{I}$ is the interference term of interest for measuring the EDM. It is due to the coherence between the $|\pm1\rangle$ states which evolves with angular frequency $\Delta_{z}$, and is proportional to $\cos^{2}\phi$ with $\phi = \Delta_{z} T$ given by Eq.~(\ref{phi}). We call the signal $S_{R}$ the Ramsey term. It depends on $\vartheta = (\omega-\Omega)T + \Phi_{\rm{rf}}$ since it arises from the interference between the $|0\rangle \leftrightarrow |c\rangle$ coherence, which evolves with angular frequency $\Omega$, and the rf field which evolves with angular frequency $\omega$. The signal $S_{C}$ is a constant background, independent of both $\phi$ and $\vartheta$. In the experiment, the two rf pulses have slightly different frequencies, but in the above model we have made them equal. Within the rotating wave approximation, the effect of the differing frequencies can simply be absorbed into the relative rf phase $\Phi_{\rm{rf}}$.

In the ideal case where the rf detunings are zero and the pulses are perfect $\pi$-pulses ($\delta_{1}=\delta_{2}=0, b_{1}\tau_{1}=b_{2}\tau_{2}=\pi$), both the constant term and the Ramsey term are zero and we are left with $S=\cos^{2}(\phi)$ as expected. Of course, the rf parameters can never be perfect and so both terms are always present. The Ramsey term is of particular concern. Because of the Stark shift of the hyperfine interval the value of $\Omega$ is sensitive to the electric field magnitude. If this magnitude changes when $E$ is reversed the phase $\vartheta$ will change, with a corresponding change in $S$. At 10\,kV\,cm$^{-1}$ the gradient of this Stark shift is $d\Omega/dE=2\pi\times 285$\,Hz/(V\,cm$^{-1}$). If the field magnitude changes by 10 parts-per-million when it reverses, the Ramsey phase $\vartheta$ changes by about $0.1$~rad. An EDM of approximately $10^{-23}$~e.cm produces the same $E$-correlated change in the interferometer phase $\phi$. Fortunately, since $S_{R}$ is proportional to $\cos(\phi)$, it does not change sign when $B$ is reversed and so is cancelled by the $B$-reversal. Still, if both the $E-$ and $B-$reversals are imperfect there will be a part of $S_{R}$ which depends on the relative directions of $E$ and $B$, just like a real EDM. We suppress this in several ways. First, the detunings and amplitudes are tuned close to their optimum values so that $S_{R}$ is minimized. Second, the phase difference $\Phi_{\rm{rf}}$ is switched between $\phi_{0} + \pi/2$ and $\phi_{0} - \pi/2$, which reverses the sign of $S_{R}$ so that it vanishes on average. Finally, on a longer timescale, $\phi_{0}$ is changed at random so that any residual $S_{R}$ averages away over the course of the experiment.

Figure \ref{Fig:Interference}(a) shows the signal measured at the probe PMT, normalized to the signal at the pump PMT. This is plotted as a function of $\phi$, which is varied using the applied magnetic field. At each point the signal is averaged over measurements made with $\Phi_{\rm{rf}} = \phi_{0} \pm \pi/2$ so that any residual Ramsey component is removed. The line is a fit to the model $S = S_{c} + S_{0}\cos^{2}(\phi - \phi_{b})$. Here, $S_{c}$ is due to background scattered laser light and un-pumped $F=0$ molecules as well as the contribution from equation (\ref{Eq:Signals}), and is approximately a third of the amplitude of the interference signal. It contributes a little to the noise in the experiment. We have included an offset phase $\phi_{b}$ in the model to account for an uncancelled background magnetic field, which here is approximately 1.5\,nT.

\subsection{Switched parameters}\label{Sec:Switches}

\begin{figure}[tb]
\centering
\includegraphics[width=0.9\columnwidth]{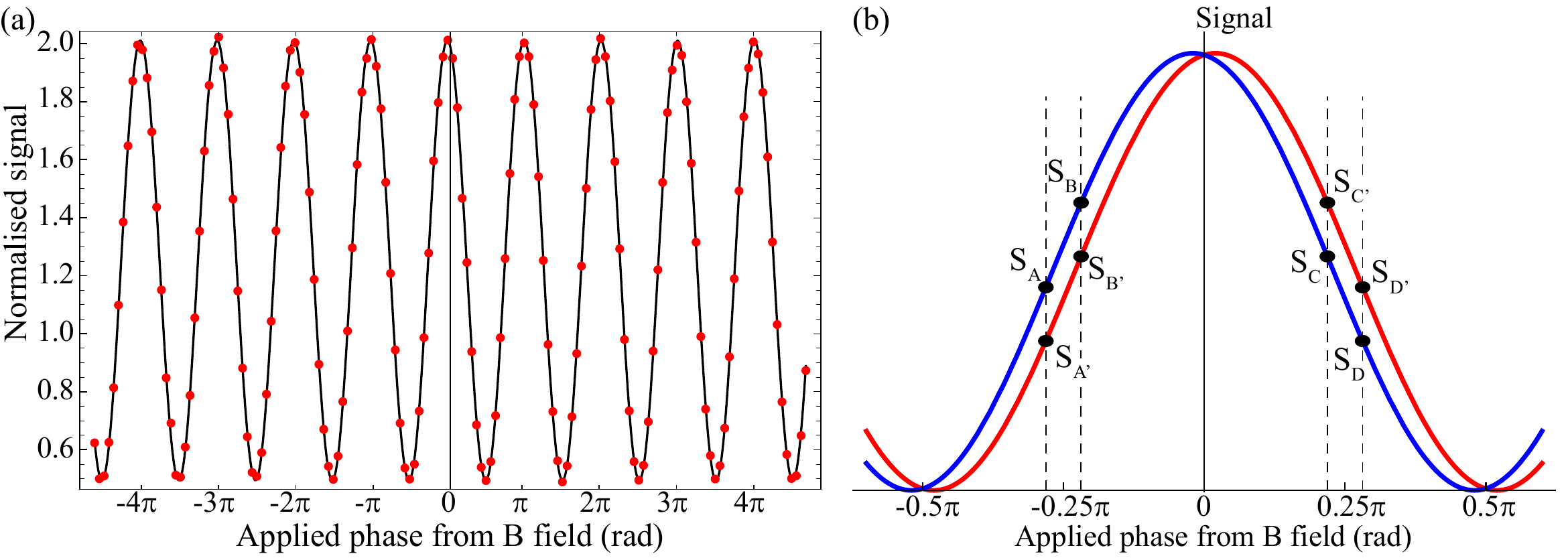}
\caption{(a) Dots show the probe signal normalized to the pump signal versus the phase applied via the B field. The line is a fit to the cosine-squared model. (b) Illustration of two interference curves, one for $E$ parallel and the other for $E$ antiparallel to $z$. The phase shift between them is due to the EDM (vastly exaggerated for clarity). Dashed lines indicate the 4 magnetic field values used in the experiment, and points show the signals obtained at these B values for the two directions of $E$.}
\label{Fig:Interference}
\end{figure}

Figure \ref{Fig:Interference}(b) illustrates the effect of a large EDM. It shows the dependence of the signal, $S$, on the applied magnetic field, $B$, for two directions of $E$, parallel and antiparallel to $z$. Reversal of $E$ produces the phase shift $\delta\phi=2 d_{e} E_{\rm{eff}}T/\hbar$ leading to a change in the signal $\delta S = (dS/d\phi)\delta\phi$. This change is maximized, and the gradient measured, by taking data at the four magnetic field values indicated by the dashed lines in figure \ref{Fig:Interference}(b). The signals obtained at these four $B$ values are $S_{A}$, $S_{B}$, $S_{C}$ and $S_{D}$ for one direction of $E$ and $S_{A'}$, $S_{B'}$, $S_{C'}$ and $S_{D'}$ for the opposite direction. The magnetic field is switched between shots and is the summed output of two switches, $B_{0}$ and $\delta B$. $B_{0}$ switches the field between $\pm 13.6$\,nT, producing phases $\phi \simeq \pm \pi/4$,  while $\delta B$ switches between $\pm 1.7$\,nT, changing the phase by approximately $\pm\pi/32$. The $\delta B$ switch calibrates the slope of the curve, and the $B_{0}$ switch ensures that spurious effects, such as a drifting background magnetic field or the Ramsey signal discussed above, are not falsely interpreted as an EDM. The electric field is switched between $\pm 10$\,kV/cm by the switch $E_0$. From the signals obtained at the eight measurement points, the EDM is

\begin{equation}\label{EDM value}
d_{\rm{e}}\!=\!\frac{g \mu_B B_{\delta B}}{E_{\rm{eff}}}\left[\frac{(S_{A}\!+\!S_{B}\!-\!S_{C}\!-\!S_{D}\!-\!S_{A'}\!-\!S_{B'}\!+\!S_{C'}\!+\!S_{D'})}{(-S_{A}\!+\!S_{B}\!+\!S_{C}\!-\!S_{D} \!-\!S_{A'}\!+\!S_{B'}\!+\!S_{C'}\!-\!S_{D'})}\right],
\end{equation}
where $B_{\delta B}=1.7$\,nT is half the change in magnetic field produced by switching $\delta B$.

In addition to $E_0$, $B_{0}$ and $\delta B$, several other parameters are switched in the experiment. The laser frequency ($\nu_{l}$) is stepped by $\pm340\,$kHz, the frequencies of the two rf pulses ($\nu_{\rm{rf1}}$ and $\nu_{\rm{rf2}}$) are independently stepped by $\pm 1.5\,$kHz, their amplitudes ($a_{\rm{rf1}}$ and $a_{\rm{rf2}}$) are independently stepped by $\pm5\%$, and the phase difference ($\Phi_{\rm{rf}}$) between them is stepped around a randomly chosen value, $\phi_{0}$, by $\pm\pi/2$. The states of these nine parameters are set between one molecular pulse and the next. There are 512 different combinations of these nine parameters, known as \emph{machine states}. We collect data in \emph{blocks} of 4096 shots, with each of the 512 machine states visited 8 times in every block. This allows us to determine how the signal correlates with any of the switched parameters, or any combination of them. The signal correlated with parameter or parameter-combination, $X$, is called a \emph{channel} and is denoted by $\left\{X\right\}$. Table \ref{Tab:Channels} gives some example channels together with their physical meaning. In equation (\ref{EDM value}), the numerator and denominator in the square brackets are $\{E_0\cdot B_{0}\}$ and $\{\delta B\}$, respectively, and so this equation can be written more succinctly as

\begin{equation}
d_{\rm{e}} = \frac{g \mu_B B_{\delta B}}{E_{\rm{eff}}} \frac{\{E_0\cdot B_{0}\}}{\{\delta B\}}.
\end{equation}

The modulations, and corresponding analysis channels, are crucial to the experimental method. They provide us with error signals that we use to servo the parameters about their optimum values. They allow us to understand how the machine behaves when the parameters are not set perfectly. They allow the EDM to be extracted robustly from the data even in the presence of a wide range of small imperfections. They also provide a method for rapidly searching for any systematic errors and provide information that can be used to correct for any systematic effects we discover.

\begin{table}[bt]
\caption{A few of the 512 analysis channels, and their relevance.}\label{Tab:Channels}
\begin{tabular}{ll}
\br
Channel & Relevance \\
\mr
$\{0\}$ & Signal averaged over all states. \\
$\{\delta B\} $ & Slope of interference curve.\\
$\{ B_0 \}$ & Value of uncancelled background magnetic field.\\
$\{\nu_{\rm{rf1}} \}$ & Average frequency detuning of rf1.\\
$\{\nu_{\rm{rf1}} \cdot B_0 \}$ & Change in interferometer phase resulting from rf1 detuning.\\
$\{\nu_{\rm{rf1}} \cdot E_0 \}$ & Change in rf1 detuning when $E_0$ is reversed, due to imperfect reversal.\\
$\{E_0 \cdot B_0 \}$ & EDM appears in this channel.\\
\br
\end{tabular}
\end{table}

For each switched parameter $X$, there is a \emph{waveform}, $W_{X}$, which specifies how $X$ is switched during a block. A waveform is a list of 4096 numbers, each having the value +1 or -1, the $i^{\rm{th}}$ element specifying the state of the switch during the  $i^{\rm{th}}$ beam shot of the block. We construct the most suitable waveforms following the methods described in \cite{switching}. Switching the parameters at high frequency ensures that low-frequency noise is eliminated from the analysis channels. Long term signal drifts are most effectively suppressed by using strongly aperiodic waveforms, i.e. waveforms made up of many different frequency components. We impose both these requirements on the all-important $E_0 \cdot B_0$ waveform. Each reversal of $E_0$ incurs a dead-time of 14.2\,s and so frequent reversal is impractical; a compromise has to be made between limiting the noise and maintaining a reasonable duty cycle. We choose to switch $E_0$ ten times in each block using a fixed waveform. The laser that provides the pump and probe beams is unstable when its frequency is modulated too rapidly so we also switch $\nu_{l}$ at low frequency using a fixed waveform. The settling time for $B$ is less than 5\,ms, much shorter than the 40\,ms time between shots, and so $B_0$ and $\delta B$ can be switched at high frequencies. The same is true for the rf parameters. There are many satisfactory waveforms and the computer randomly chooses new ones for all parameters except $E_0$ and $\nu_{l}$ at the beginning of every block. This randomization prevents specific frequency components of the background from leaking into the analysis channels the same way in every block. A random selection of the waveforms are inverted so that blocks may start with parameters either in the high or the low state. To ensure that none of the machine states are missed out, the 512 waveforms corresponding to the 512 analysis channels must all be distinct. The computer checks this is the case and, if not, chooses new waveforms.

\subsection{Data acquisition}\label{Sec:DataAcquisition}

For each shot of the experiment, data is acquired from the probe PMT, the pump PMT, a magnetometer situated between the two magnetic shields, three other magnetometers placed strategically around the laboratory, two ammeters that monitor the currents flowing to the electric field plates, and two dummy inputs (a battery and a short-circuit). The states of the 9 switched parameters are set between one shot and the next according to the waveforms discussed above. During a block, shots are fired with a repetition rate of 25\,Hz. The source continues to run during the dead-time when $E_0$ is being switched, but no data is taken during this time.

At the end of each block, error signals are derived from the channel values and used to make small adjustments to the parameters. For example, the bias magnetic field, which is used to cancel any background field, is adjusted between blocks according to the value of $\{B_{0}\}/\{\delta B\}$. The rf frequencies and amplitudes and the laser frequency are also automatically adjusted between blocks according to the values of $\{\nu_{\rm{rf1}}\cdot \delta B\}/\{\delta B\}$ etc. The aim of these servo loops is to bring these channel values to zero. The pump and probe polarizer angles and the rf phase $\phi_{0}$ are set to new, randomly chosen values, and new waveforms are selected before the next block of data is acquired. Including the dead-time, each block takes 6 minutes to acquire.

As well as the automated switching of the parameters, we also occasionally make \emph{manual reversals}. The high-voltage connections are swapped to reverse $\vec{E}$, the connections to the magnetic field coil are swapped to reverse $\vec{B}$ and the rf cables are swapped to reverse the direction of rf propagation along the field plates. The manual reversals are usually made after obtaining 50\,--\,100 consecutive blocks. They are valuable for identifying and eliminating systematic effects. We make sure to obtain roughly equal numbers of blocks in all 8 of these \emph{manual configurations}. When averaging channels containing $E_0$ or $B_{0}$ over a set of blocks, we will usually want to include a sign to account for the manual configuration of these fields in each block. We use the symbols $\cal{E}$ and $\cal{B}$, which take the values $\pm 1$, for this purpose. Similarly, to denote the direction of rf propagation we use the symbol $\cal{\nu}$.

\section{Analysis of the data}

The main dataset for the present measurement contains 6,194 blocks, or equivalently 25,370,624 shots. We now describe how this data is analyzed. Throughout the analysis the EDM values themselves were concealed by adding a fixed unknown offset which was only removed once the analysis was complete. This offset was a number chosen at random by the computer from a normal distribution with zero mean and a standard deviation of $5 \times 10^{-27} e$\,cm.

\subsection{Calculating channel values}

We first calculate the 512 channel values for every block. The value of channel $X$ is given by
\begin{equation}\label{Eq:ChannelValues}
\{X\}(t)=\frac{1}{N}\sum_{i=1}^{N} W_{X}(i)\,\frac{S_{i}(t)}{{\cal N}_{i}}\\
\end{equation}
where $N=4096$ is the number of shots in the block, $W_{X}$ is the waveform for $X$, $S_{i}(t)$ is the probe PMT signal at time $t$ for the $i^{\rm{th}}$ shot and ${\cal N}_{i}$ is the integral of the pump PMT signal for the $i$'th shot. Since the probe signal is a function of time, so too are the channels. We have indicated this explicitly here, though we will often simply use $\{X\}$, the time-dependence being understood. Note that while the time-dependence of the probe signal has been retained, this signal is normalized to the time-integrated pump PMT signal which simply measures the number of molecules produced in the shot.

The value of $\{X\}(t)$ integrated over some chosen time window is denoted $\langle \{X\} \rangle$ and is given by
\begin{equation}\label{Eq:IntegratedChannelValues}
\langle \{X\} \rangle = \int^{t_f}_{t_i} \{X\}(t) \rm{dt}.
\end{equation}
In our analysis we have used the values of $t_i$ and $t_f$ shown by the dashed lines in figure \ref{fig:tof}. This choice minimizes the statistical uncertainty of the measurement. A narrower time window decreases the number of molecules used in the experiment, which increases the uncertainty. A wider time window corresponds to a longer bunch of molecules, increasing the inhomogeneity of the static and rf fields sampled by the molecules and thereby also increasing the measurement uncertainty.

We are often interested in products or ratios of channels, such as $\{X\}\{Y\}/\{Z\}$. The time-integrated value of this quantity is $\langle\{X\}\{Y\}/\{Z\}\rangle$, the integration being done at the end\footnote{as opposed to $\langle\{X\}\rangle \langle\{Y\}\rangle / \langle\{Z\}\rangle$}. Note carefully that $\{X\}\{Y\}$ is an entirely different quantity from $\{X \cdot Y\}$.

\subsection{Accounting for non\,-\,ideal changes in lineshape}\label{Sec:Correction}

In section \ref{Sec:Switches} we showed that, for an ideal experiment, the EDM is proportional to $\{E_0 \cdot B_0 \}/\{\delta B\}$. Now we consider how to extract the EDM from the data when the experiment is not ideal. In particular, we allow for an uncancelled background magnetic field $B_{\rm{back}}$, and we allow the amplitude of the interference curve to change by a small amount $2a$ when $E_0$ is reversed and by a small amount $2b$ when $B_{0}$ is reversed \footnote{The $a$ and $b$ used here are not the same as those in equation (\ref{Eq:ab})}. In this model, the signal is $S(\widehat{E_0},\widehat{B_0},\widehat{\delta B}) = A f(\Theta)$ where $\widehat{X}$ denotes the state ($\pm 1$) of switch $X$, $A$ is an amplitude given by
\begin{equation}
A = 1 + a \widehat{E_0} + b \widehat{B_{0}},
\end{equation}
and $f$ is an arbitrary function of the interferometer phase $\Theta$, with
\begin{equation}
\Theta = \phi_{\rm{back}} + \phi_{B_0} \widehat{B_0} + \phi_{\delta B} \widehat{\delta B}\,\widehat{B_0} + \phi_{\rm{EDM}} \widehat{E_0}.
\end{equation}
Here, $\phi_{\rm{back}}$, $\phi_{B_0}$, and $\phi_{\delta B}$ are the magnetic phases due to $B_{\rm{back}}$, $B_0$ and $\delta B$, and $\phi_{\rm{EDM}}$ is the phase due to the EDM.

Provided the background magnetic field is small and $\phi_{B_0}$ is chosen appropriately, the interference curve is very nearly linear at all points of interest and we can expand $f(\Theta)$ about the points $\phi_{\rm{back}} \pm \phi_{B_0}$, retaining only the constant and linear terms. Then, writing out the signals in all the switch states, and using the notation $f_{\pm}=f(\phi_{\rm{back}} \pm \phi_{B_0})$, $f'_{\pm}=\frac{df}{d\Theta}|_{\phi_{\rm{back}} \pm \phi_{B_0}}$, we obtain the following expressions for the channels:
\begin{eqnarray}
&\{0\}\!=\!\frac{1}{2}\left(f_+\!+\!f_-\right)\!+\!\frac{1}{2}b\left(f_+\!-\!f_-\right)\!\simeq\!\frac{1}{2}\left(f_+\!+\!f_-\right),\nonumber\\
&\{\delta B\}\!=\!\frac{1}{2}\left[\left(f_+'\!-\!f_-'\right)\!+\!b\left(f_+'\!+\!f_-'\right)\right]\phi _{\delta B}\!\simeq\!\frac{1}{2}\left(f_+'\!-\!f_-'\right)\phi _{\delta B},\nonumber\\
&\{E_0\}\!=\!\frac{1}{2}a\left(f_+\!+\!f_-\right),\nonumber\\
&\{B_0\}\!=\!\frac{1}{2}\left(f_+\!-\!f_-\right)\!+\!\frac{1}{2}b\left(f_+\!+\!f_-\right),\nonumber\\
&\{E_0 \cdot B_0\}\!=\!\frac{1}{2}a\left(f_+\!-\!f_-\right)+\frac{1}{2}\left(f_+'\!-\!f_-'\right)\phi _{\rm{EDM}},\nonumber\\
&\{E_0 \cdot \delta B\}\!=\!\frac{1}{2}a\left(f_+'\!-\!f_-'\right)\phi _{\delta B},\nonumber\\
&\{B_0 \cdot \delta B\}\!=\!\frac{1}{2}\left(f_+'\!+\!f_-'\right)\phi _{\delta B}\!+\!\frac{1}{2}b\left(f_+'\!-\!f_-'\right)\phi _{\delta B},\nonumber\\
&\{E_0 \cdot B_0 \cdot \delta B \}=\frac{1}{2}a\left(f_+'\!+\!f_-'\right)\phi _{\delta B}.\nonumber
\end{eqnarray}
In deriving these expressions, we have neglected terms of the form $\gamma \phi_{\rm{EDM}}$ where $\gamma \ll 1$. Provided that $\phi_{\rm{back}} \ll \phi_{B_0}$ and that $E_0$ and $B_0$ reversals are not too imperfect, $a$, $b$, $(f_+\!-\!f_-)$ and $(f_+'\!+\!f_-')$ are all small quantities. In the expression for $\{0\}$ the second term is the product of two small quantities whereas the first is of order 1, and this leads to the approximate result we have given. The same applies to $\{\delta B\}$.

As expected, $\{E_0 \cdot B_0\}$ is the channel that is sensitive to the EDM, but this channel now contains another contribution which is due to the change in amplitude upon $E_0$ reversal (proportional to $a$) multiplying the non-zero $B_{\rm{back}}$ (proportional to $(f_+\!-f_-)$). This extra contribution can be cancelled using the product of the two channels that tell us about these two imperfections, namely $\{E_0 \cdot \delta B\}$ and $\{B_0\}$. This cancellation would work perfectly if the $\{B_0\}$ channel only measured $B_{\rm{back}}$, but the amplitude change due to $\{B_0\}$ reversal (proportional to $b$) also contributes to this channel. Once again, this extra term can be cancelled using a combination of the other channels. Thus, we obtain the following expression for the EDM which is valid in the presence of the imperfections:

\begin{eqnarray}\label{Eq:LineshapeCorrectedEDM}
\fl d_e\!=\!\frac{g \mu_{B} B_{\delta B}}{E_{\rm{eff}}}{\cal E}{\cal B}
\left[\frac{\{E_0 \cdot B_0\}}{\{\delta B\}} -\frac{\{B_0\}\{E_0 \cdot \delta B\}}{\{\delta B\}\{\delta B\}} + \right.\\ \nonumber
\left. \frac{\{0\}}{\{\delta B\}}\left(\frac{\{B_0 \cdot \delta B\}\{E_0 \cdot \delta B\}}{\{\delta B\}\{\delta B\}}\!-\!\frac{\{E_0 \cdot B_0 \cdot \delta B\}}{\{\delta B\}}\right)\right].
\end{eqnarray}

Here, we have included the signing according to the manual configuration of the machine, denoted by ${\cal E}$ and ${\cal B}$. We may regard the terms beyond $\{E_0 \cdot B_0\}$ as lineshape correction terms. Higher order correction terms coming from the next order of the Taylor expansion of $f(\Theta)$ and from the small terms we neglected in $\{0\}$ and $\{\delta B\}$ are much smaller than the terms we have written down. For the measurement presented in this paper, the leading-order correction terms (the ones in equation (\ref{Eq:LineshapeCorrectedEDM})) are 4 times smaller than the statistical uncertainty in $d_e$, and so higher-order correction terms are negligible. We have also considered other models for how the amplitude $A$ might change with the switches, for example one where the height of the interference curve changes but the background does not. Such changes to the model change the pre-factor $\{0\}/\{\delta B\}$ in the last term of equation (\ref{Eq:LineshapeCorrectedEDM}), but since this term was negligible in the experiment the details are unimportant. We might also wonder about lineshape-changing effects arising from imperfections in the other switched parameters. In \ref{AppA} we use a generalization of the procedure outlined here to derive a more general expression for $d_e$ which accounts for these additional imperfections. None of the extra terms were important for the present measurement.

In section \ref{Sec:rf phase correction}, we describe the measurement of an interferometer phase arising from a detuning of rf1(2). This phase is measured by the $\{\nu_{\rm{rf1(2)}} \cdot B_0 \}$ channel. In fact, to make sure that this phase is measured correctly in the presence of non-ideal changes to the lineshape, we correct $\{\nu_{\rm{rf1(2)}} \cdot B_0 \}$ using the same correction terms as in equation (\ref{Eq:LineshapeCorrectedEDM}), but with $E_0$ replaced by $\nu_{\rm{rf1(2)}}$.

\subsection{Mean and statistical uncertainty}\label{EDM mean and error}

\begin{figure}[t]
\begin{center}
\includegraphics[width=0.6\columnwidth]{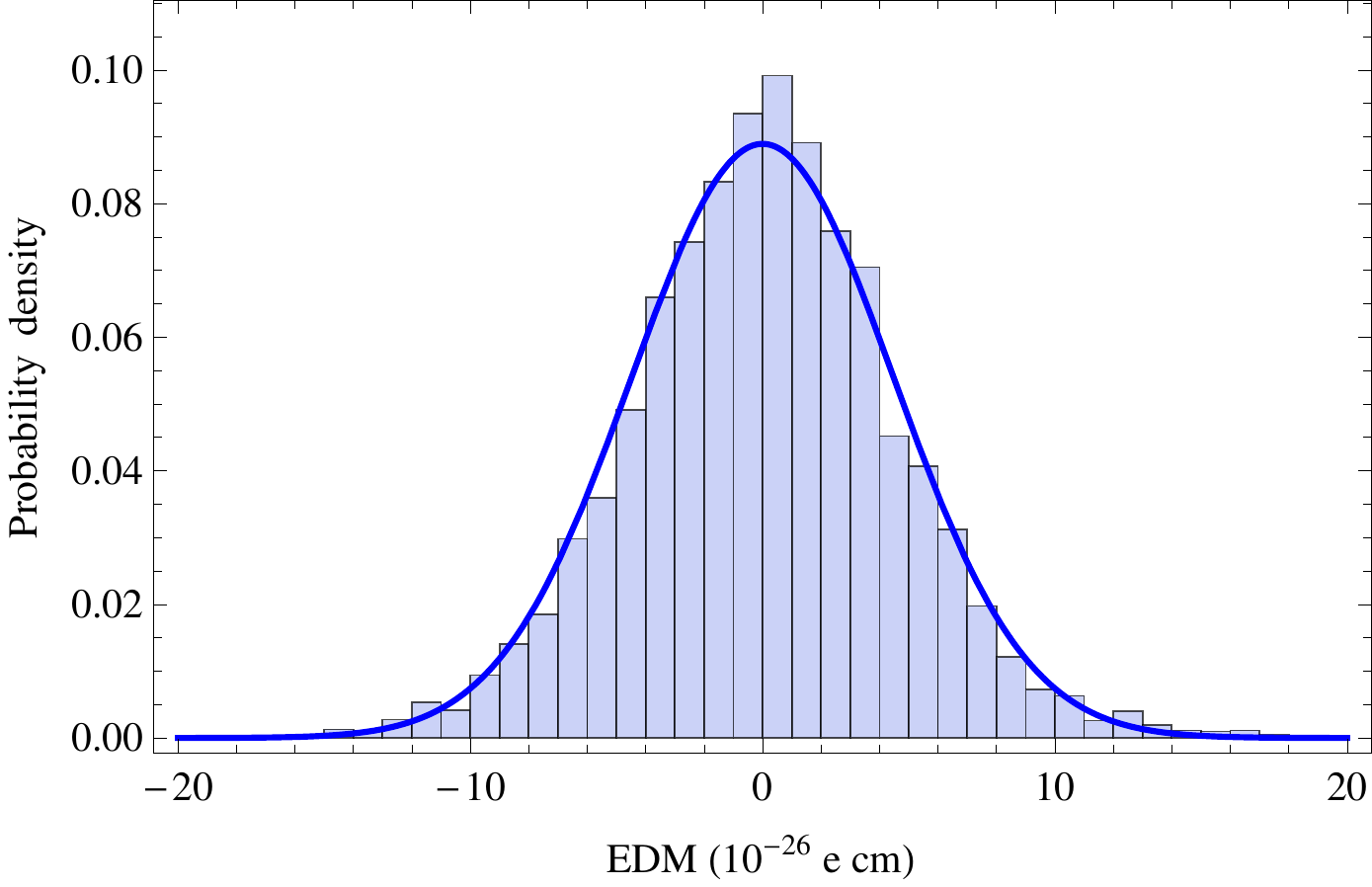}
\caption{\label{Fig:edmHistogram} Histogram of the EDM values measured in each block. The line is a normal distribution with a mean of zero and a standard deviation equal to the average statistical uncertainty per block.}
\end{center}
\end{figure}

Figure \ref{Fig:edmHistogram} shows the distribution of the EDM values, determined using equation (\ref{Eq:LineshapeCorrectedEDM}). The line is a normal distribution centred at zero and with a standard deviation equal to the average statistical uncertainty per block. We see that our distribution deviates a little from a normal distribution, there being a small excess of points near the centre and also in the wings of the distribution (though the latter is not visible in figure \ref{Fig:edmHistogram}). This deviation is mainly due to variations in the sensitivity of the experiment.

For non\,-\,normal distributions the mean and standard error are not robust measures. For the central values, we use instead the 5\% trimmed mean, a simple and robust statistic that drops the smallest and largest 5\% of the values. To find the statistical uncertainties we use the bootstrap method~\cite{Robust Statistics, Bootstrap methods for standard errors}. From the real experimental dataset, containing $n$ measurements of the quantity $v$, we create $m$ (usually 5000) synthetic datasets, each generated by drawing a random selection of $n$ points from the real dataset. Points can be drawn more than once, so the synthetic datasets are all different. We then calculate the trimmed mean of each dataset by removing the smallest and largest 5\,\% of the values and then taking the mean. We use the cumulative distribution function of these trimmed mean values, ${\rm CDF}(v)$, to calculate symmetric confidence intervals. If the probability of obtaining a result within the interval is $c$, the boundaries of the interval are at $v_{\pm}$ where ${\rm CDF}(v_{\pm})=0.5\pm c/2$. For this confidence interval, the central value is $\bar{v}=\frac{1}{2}(v_+ + v_-)$ and the statistical uncertainty is $\sigma_{v}=\frac{1}{2}(v_+ - v_-)$. We use these methods to find the central value and uncertainty of the EDM and all other quantities of interest.

Before we can give the result for the central value of the EDM and the statistical uncertainty (section \ref{Sec:Result}), we must consider some systematic errors in the experiment and the corrections applied to account for them.

\subsection{rf detuning correction}\label{Sec:rf phase correction}

The analysis of the data revealed that a detuning of either rf frequency changes the phase of the interferometer. This phase change produces a signal in the $\{\nu_{\rm{rf1(2)}} \cdot B_0 \}$ channel. Specifically, the rate of change of phase with rf detuning is given by
\begin{equation}
\frac{d\phi}{d \nu_{\rm{rf1(2)}}} = \frac{\{\nu_{\rm{rf1(2)}} \cdot B_0 \}}{\{\delta B \}} \frac{\delta \phi_{\delta B}}{\delta \nu_{\rm{rf1(2)}}}
\end{equation}
where $\delta \phi_{\delta B} \simeq \pi/32$ is the phase change due to $\delta B$ and $2 \delta \nu_{\rm{rf1(2)}} \simeq 3$\,kHz is the full size of the rf1(2) frequency step. We find that this phase changes differently when detuning rf1 and rf2, and that it depends on the direction of propagation of the rf field. On detuning of rf1 the derivative is $316 \pm 8$\,nrad\,Hz$^{-1}$ for downward propagation, and $261 \pm 8$\,nrad\,Hz$^{-1}$ for upward propagation. For rf2 the equivalent values are $-139 \pm 7$\,nrad\,Hz$^{-1}$ and $-42 \pm 8$\,nrad\,Hz$^{-1}$. In our numerical modelling of the experiment, we have not found a way to produce such large phases. Our model includes imperfect settings of the rf frequency, amplitude and polarization, the standing-wave component of the rf field as measured in \cite{Hudson(1)07}, and realistic changes in the magnetic field, the Stark-shifted hyperfine frequency and the rf field as the molecules move during the rf pulse. The model does not yet include the rf electric field or the possibility of a coherence set-up by the optical pumping process.

\begin{figure}[t]
\begin{center}
\includegraphics[width=\columnwidth]{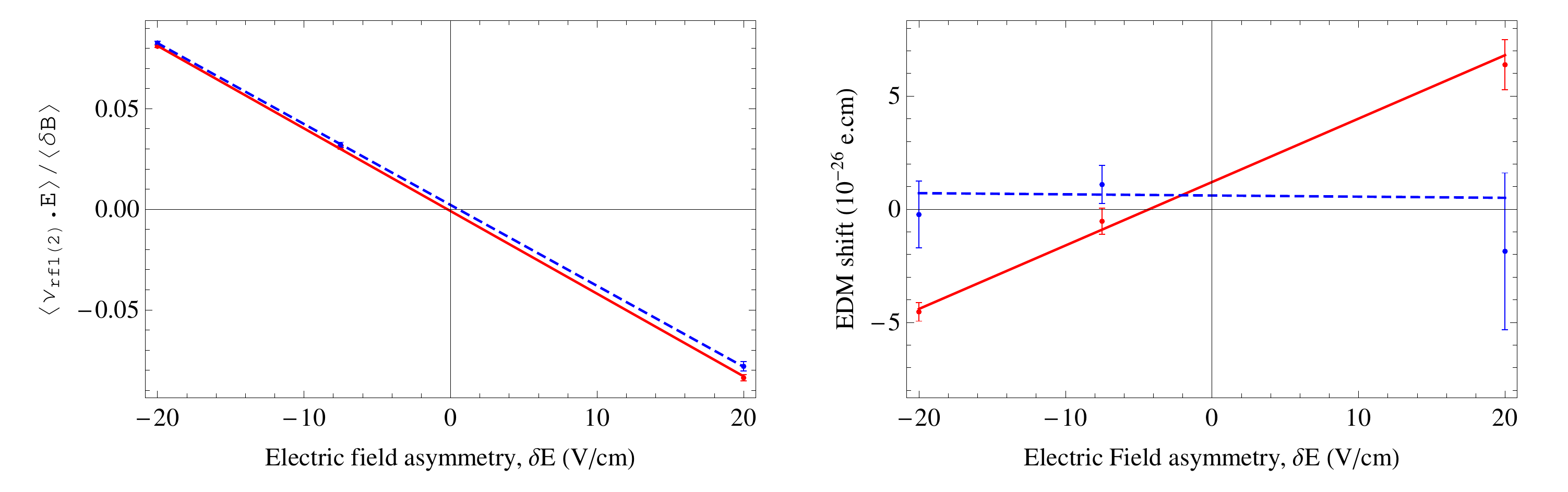}
\caption{\label{Fig:rfPhaseCorrection} (a) Calibration of $\frac{\{\nu_{\rm{rf1(2)}} \cdot E_0 \}}{\{\delta B \}}$ versus $\delta E$. Red is for rf1 and blue is for rf2. (b) Systematic EDM shift versus $\delta E$. Red points: before correction. Solid red line: straight line fit to these points. Blue points: after correction using (\ref{Eq:rfPhaseCorrection}). Dashed blue line: straight line fit to these points.}
\end{center}
\end{figure}

Whatever their cause, these rf detuning induced phases result in a systematic EDM shift when combined with an electric field reversal that is imperfect. Suppose that the magnitude of the electric field changes by $\delta E$ when $E_0$ is switched. Due to the Stark shift of the hyperfine interval, there will be a corresponding change in the rf detunings of $s\,\delta E$ where $s=285$\,Hz/(V cm$^{-1}$) is the gradient of the Stark shift at the operating electric field. This $E_0$-correlated change in rf detuning produces a signal in the $\{\nu_{\rm{rf1(2)}} \cdot E_0 \}$ channel. We assume the linear relationship
\begin{equation}\label{Eq:deltaE}
\frac{\{\nu_{\rm{rf1(2)}} \cdot E_0 \}}{\{\delta B\}} = \beta_{1(2)} \delta E
\end{equation}
and determine the proportionality constants $\beta_{1(2)}$ in a separate experiment where we deliberately applied large values of $\delta E$. The result of this calibration is shown in figure \ref{Fig:rfPhaseCorrection}(a). We find the two proportionality constants to be equal within their uncertainties. The size of the actual electric field asymmetry for the main data is discussed in section \ref{Sec:ElectricFieldAsymmetry}.

Together, the imperfect $E$-reversal and the detuning-induced phases generate an interferometer phase change that is correlated with $E_0$, and hence a systematic error. The resulting false EDM, with signing due to the manual configuration included, is
\begin{equation}\label{Eq:rfPhaseCorrection}
d_{\rm{rf1(2)}} = \frac{g \mu_{B} \delta B}{E_{\rm{eff}}} \frac{1}{\delta \nu_{\rm{rf1(2)}}} \frac{s}{2\beta} \frac{\{\nu_{\rm{rf1(2)}} \cdot B_0  \}{\cal B}}{\{\delta B \}} \frac{\{\nu_{\rm{rf1(2)}} \cdot E_0 \}{\cal E}}{\{\delta B \}}.
\end{equation}
For each block of data, we apply a correction to the EDM by subtracting $d_{\rm{rf1(2)}}$ given by (\ref{Eq:rfPhaseCorrection}) using the values of $\{\nu_{\rm{rf1(2)}} \cdot E_0 \}{\cal E} / \{\delta B \}$ and $\{\nu_{\rm{rf1(2)}} \cdot B_0 \} {\cal B}/ \{\delta B \}$ measured in each block. We have tested that this correction works by taking EDM data with large electric field asymmetries deliberately applied. The results are shown in figure \ref{Fig:rfPhaseCorrection}(b) where the red points are the uncorrected EDM values and the blue points are the EDM values obtained after subtracting the correction given by (\ref{Eq:rfPhaseCorrection}). While there is a clear dependence on $\delta E$ prior to correction, this dependence vanishes once the correction is applied.

Applying the correction to the main dataset, and averaging over all blocks, the rf1 phase correction is $(5.0 \pm 0.9) \times 10^{-28}$\,e.cm and the rf2 phase correction is $(0.5 \pm 0.7) \times 10^{-28}$\,e.cm. These averages are given here for reference though we do not use them in the analysis where the corrections are made to each block and the average of these corrected EDMs is taken.

\subsection{Magnetic field correction}\label{Sec:magCorrection}

A major concern in the experiment is that the background magnetic field may change when the electric field is reversed. Random changes will increase the spread of the measurements and so increase the statistical uncertainty, while a systematic correlation will produce a systematic shift to the measured EDM. As we will see, in our present measurement there is no overall systematic shift due to such an $E_0$-correlated magnetic field, but magnetic noise does contribute (a little) to the statistical uncertainty.

\begin{figure}[t]
\begin{center}
\includegraphics[width=0.6\columnwidth]{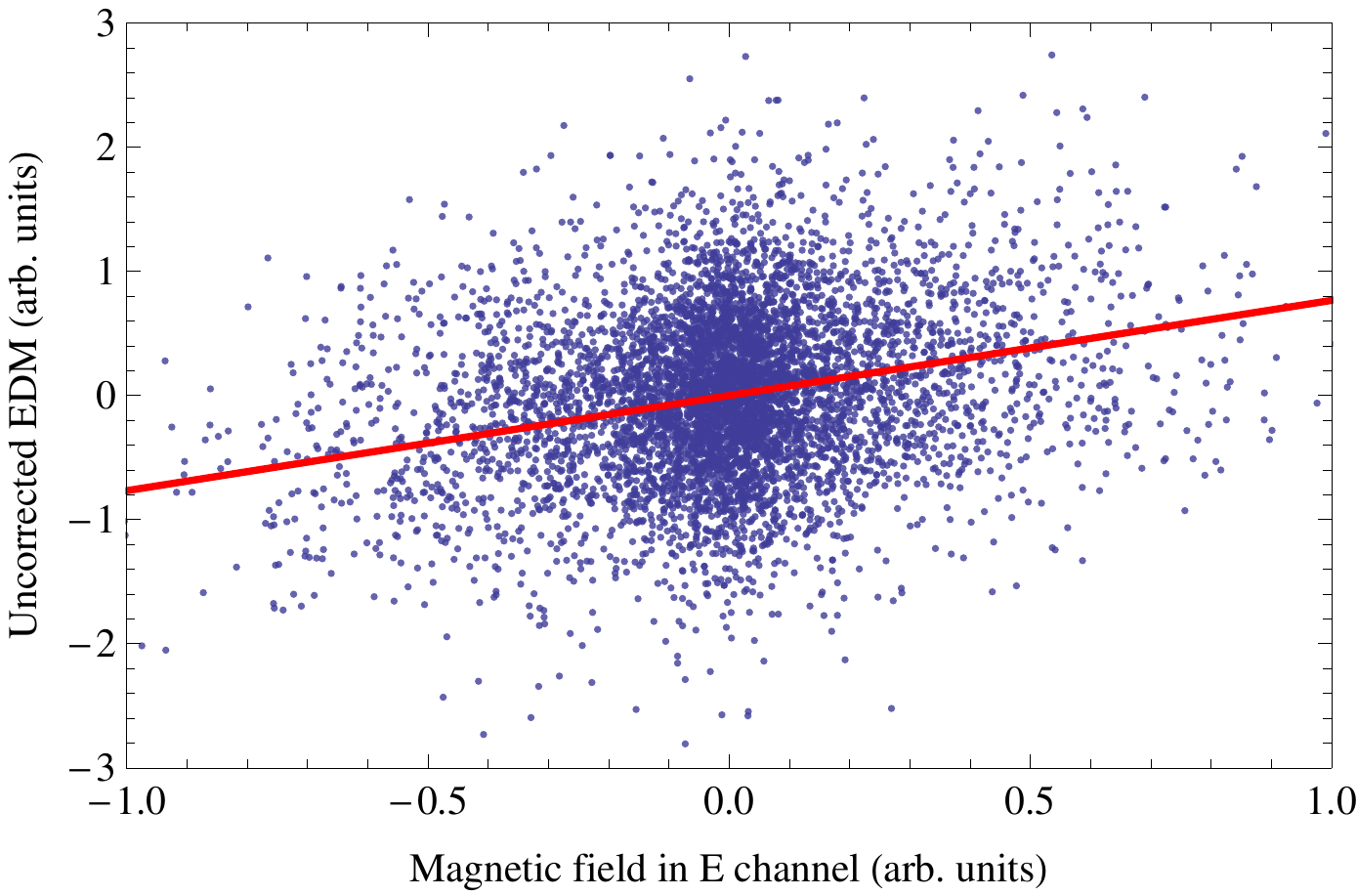}
\caption{\label{Fig:magneticCorrelation} EDM versus the part of the magnetometer signal that correlates with the $E_0$-reversal. The line is a straight line fit to the points and shows that the two quantities are correlated. The slope of this fit is used to correct the EDM data.}
\end{center}
\end{figure}

The two layers of magnetic shielding reduce the background magnetic field in the interaction region. A fluxgate magnetometer situated between these two shields samples the field in the $z$-direction during each shot of the experiment. The data from this magnetometer is analyzed in the same way as the main data from the probe PMT. Of particular interest is the magnetometer signal in the $\{E_0\}$ channel, $\{E_0\}_{\rm{mag}}$, since this measures the change in the magnetic field that correlates with the $E_0$-reversal. Averaged over the dataset, its value is consistent with zero. Figure \ref{Fig:magneticCorrelation} shows, for each block, the EDM versus $\{E_0\}_{\rm{mag}}$. There is a statistically significant correlation between these two quantities, as we would expect since they are both sensitive to the component of the magnetic field noise that is correlated with the switching of $E_0$. The line in figure \ref{Fig:magneticCorrelation} is a straight line fit to the data, which determines the gradient $\alpha$. We note that the uncertainties in the two variables are comparable, and that ordinary least squares regression underestimates the gradient in this case. In our fit, we have corrected for this underestimate using a `reliability ratio' \cite{Carroll(1)95}, which for this data is 0.83.

From the EDMs measured in each block we subtract the quantity $\alpha \{E_0\}_{\rm{mag}}$. This correction reduces the contribution of magnetic noise to the statistical uncertainty of the measurement, though it introduces the intrinsic noise of the magnetometer into the measurement. Overall, the statistical uncertainty is reduced by 3.5\%. The mean size of the correction is $(-0.3 \pm 1.7) \times 10^{-28}$\,e.cm.

\section{Systematic tests}
\label{Sec:SystematicTests}

A number of auxiliary measurements were made to search for possible systematic errors. These are measurements under exaggerated non-ideal conditions, used to set limits on systematic shifts. All systematic shifts to the EDM are due to changes that occur in the experiment when the electric field is switched. We start by discussing changes that may occur to the magnitude or direction of the electric field.

\subsection{Change of electric field magnitude on reversal}\label{Sec:ElectricFieldAsymmetry}

When the electric field is switched, relays reverse the connections between the high voltage supplies and the field plates. Ideally, this would result in an exact field reversal, but in practice the field magnitude changes slightly. In section \ref{Sec:rf phase correction} we saw an example of how an imperfect field reversal results in a systematic error. The mechanism discussed there is just one of several possibilities. For example, the electric field may leak into the pump or probe region and effect the pumping or detection efficiency, so that if the field magnitude changes when the field reverses there will be an $E_0$-correlated change in the detected signal. Another possibility, discussed in section \ref{Sec:eDependentBSensitivity}, is that the sensitivity of the molecule to magnetic fields may depend on the electric field magnitude. There are no doubt other mechanisms that we have not thought of, so in this section we describe our approach to empirically constrain the effect of \emph{all} possible mechanisms.

\begin{figure}[t]
\begin{center}
\includegraphics[width=0.6\columnwidth]{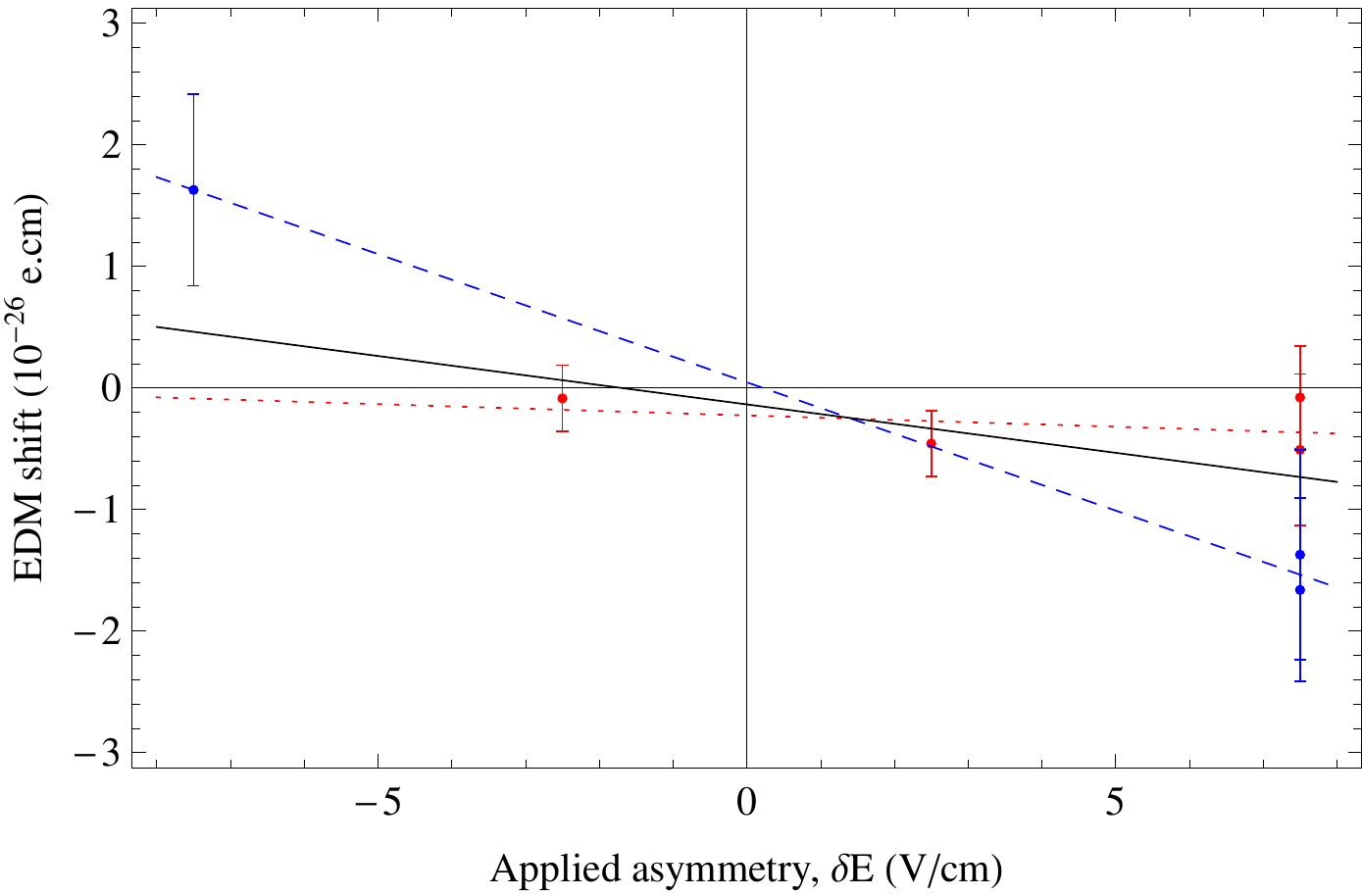}
\caption{\label{Fig:electricFieldAsymmetry} Measured EDM versus applied electric field asymmetry, $\delta E$. For this data there is also an offset voltage of $\bar{V} = 102.5$\,V (see section \ref{Sec:offsetVoltage}). Red points: rf travels upwards. Blue points: rf travels downwards. Straight line fits are shown to the whole set of data (solid black line), to the red points only (dotted red line) and to the blue points only (dashed blue line).}
\end{center}
\end{figure}

To measure the effect of imperfect field reversal we made measurements of the EDM with large electric field asymmetries applied. Upon switching $E_0$ we changed the electric field magnitude by $\delta E$ by changing the voltage of one of the power supplies. For each value of $\delta E$, the apparent EDM was determined using the same analysis procedure as for the main dataset. This procedure includes the rf phase correction described in section \ref{Sec:rf phase correction} so that this known effect due to $\delta E$ is removed from the data. The measurements then expose any other possible systematic effects related to $\delta E$. Measurements of this kind are shown in figure \ref{Fig:electricFieldAsymmetry}. The plot includes data taken for both directions of rf propagation and we distinguish these since there appears to be a small difference between them. At $\delta E = 7.5$\,V/cm we took data in both manual $E$ configurations and we show these measurements as separate data points though we will not distinguish them since their results are consistent with one another. The solid black line in figure \ref{Fig:electricFieldAsymmetry} is a straight line fit to all of the data points plotted and has a gradient of $(-8.0 \pm 3.6)\times 10^{-28}\,e\,\rm{cm}/(\rm{V}\,\rm{cm}^{-1})$. Treating the data for the two directions of rf propagation separately, we obtain the two straight line fits shown by the red and blue lines in figure \ref{Fig:electricFieldAsymmetry}. For the upward propagation data the gradient is $(-1.9 \pm 4.4)\times 10^{-28}\,e\,\rm{cm}/(\rm{V}\,\rm{cm}^{-1})$ and so shows no indication of any dependence on $\delta E$. For the downward propagation data the gradient is $(-21.1 \pm 6.5)\times 10^{-28}\,e\,\rm{cm}/(\rm{V}\,\rm{cm}^{-1})$ which differs from zero by 3.2 standard deviations. Note that for the data shown in figure \ref{Fig:electricFieldAsymmetry} there is also an offset voltage of $\bar{V} = 102.5$\,V (its meaning is explained in section \ref{Sec:offsetVoltage}). Equivalent data for $\bar{V} = 0$\,V are shown by the blue points in figure \ref{Fig:rfPhaseCorrection}(b) which shows no dependence on $\delta E$  - the gradient is  $(-0.5 \pm 8.8)\times 10^{-28}\,e\,\rm{cm}/(\rm{V}\,\rm{cm}^{-1})$. We take the view that there is no strong evidence for any residual systematic shifts that depend on $\delta E$, and so do not make any further corrections to the main EDM dataset. However, since we find a hint of a dependence on $\delta E$ when $\bar{V} = 102.5$\,V and when the rf propagates downwards, we suppose, quite conservatively, that this same effect might also apply to the main dataset (which has $\bar{V}=0$\,V). Since only half the dataset has the rf propagating downwards, we use half the gradient found for the downward propagation direction, $-11 \times 10^{-28}\,e\,\rm{cm}/(\rm{V}\,\rm{cm}^{-1})$, to determine the related systematic uncertainty.

This gradient must be multiplied by an estimate of the actual electric field asymmetry in the main dataset. This we obtain using equation (\ref{Eq:deltaE}) and the measured value of $\{\nu_{\rm{rf1(2)}} \cdot E_0 \}/\{\delta B\}$. It is useful to distinguish two sources of asymmetry, external and internal to the machine. A change in a power supply voltage when the relays are switched is an example of an external source of $\delta E$, whereas patch potentials on the electric field plates are an example of an internal source. The external asymmetry can be isolated by averaging $\{\nu_{\rm{rf1(2)}} \cdot E_0 \}$ over the two manual $E$ configurations. Then, the asymmetries measured in the two rf regions are consistent with one another, as expected for an external source, and we find $\delta E_{\rm{ext}} = 0.463 \pm 0.015$\,V\,cm$^{-1}$. The internal asymmetry can be isolated by signing $\{\nu_{\rm{rf1(2)}} \cdot E_0 \}$ according to the manual $E$ configuration (i.e. by ${\cal E}$) and then averaging. We find different asymmetries in the two rf regions, $\delta E_{\rm{int}} = 0.21 \pm 0.02$\,V\,cm$^{-1}$ in the first rf region and $\delta E_{\rm{int}} = -0.21 \pm 0.02$\,V\,cm$^{-1}$ in the second rf region. That they are different suggests the presence of local patch potentials on the electric field plates. Because reversal of the manual $E$ configuration changes the sign of the EDM shift due to $\delta E_{\rm{ext}}$ but does not change the sign of the shift from $\delta E_{\rm{int}}$, only $\delta E_{\rm{int}}$ contributes to the systematic uncertainty. We take the asymmetries measured in the two rf regions to be typical, but since the rf regions occupy a substantial fraction of the whole interaction region and the asymmetries are equal and opposite in these two regions we expect some cancellation of any asymmetry-induced effect when averaged over the interaction region. So we take a characteristic value of $|\delta E| = 0.1$\,V\,cm$^{-1}$ giving a systematic uncertainty due to uncorrected $\delta E$ effects of $1.1 \times 10^{-28}\,e\,\rm{cm}$.

\subsection{Electric field ground offset}\label{Sec:offsetVoltage}

Now we discuss a second type of imperfect electric field reversal. We aim to charge the field plates to equal and opposite potentials, $\pm V$. However, imperfect setting of the power supplies will result in the plates being asymmetrically charged with respect to ground, to potentials $\pm V + \bar{V}$. We call the mean potential on the plates, $\bar{V}$, the offset voltage. Due to the presence of grounded support structures, and the magnetic shield, a non-zero $\bar{V}$ results in a change in the field distribution when the relays are switched. This could lead to a systematic EDM shift. Let us give some examples. The molecules are only sensitive to the magnetic field component parallel to the electric field \cite{Hudson(1)02}, and so the interferometer phase accrued from the magnetic interaction is proportional to the magnetic field projected along the electric field direction, integrated along the molecular trajectory. If the electric field direction does not reverse perfectly this magnetic phase can change. Similarly, a change in the distribution of the electric field can result in a change in the geometric phase (see section \ref{Sec:geometricPhase}) when $E$ is reversed. Another possible effect is due to the rf field whose direction rotates as it enters and exits the transmission line. The phase imprinted by the rf pulses depends on the projection of the rf magnetic field vector into the plane perpendicular to the electric field. A change in direction of the electric field would result in a change in this phase. Detailed numerical modelling suggests that all of the above effects are negligibly small in the experiment. This is because the rf pulses are applied when the molecules are far away from the edges of the plates, and here the electric field is little affected by the relatively distant grounded surfaces. Conversely, a non-zero $\bar{V}$ can cause the fringe fields near the edges of the plates to change significantly on $E$-reversal. Leakage of these fringe fields into the optical pumping or detection regions could change the output of the interferometer. For example, imperfect optical pumping might produce a coherence in the $F=1$ level, and when this is combined with imperfect rf pulses and the $E$-correlated change in the fringe fields, there could be a systematic EDM shift. We have not modelled this type of effect.

To investigate these and any other possible effects empirically we acquired data with large applied offset voltages of $-1000.5$\,V, $+102.5$\,V and $+1015$\,V. Each of these datasets was analysed to reveal the apparent EDM, using the same procedure as used for the main dataset. We found two distinct effects in these datasets. Here, we describe these effects and analyze the impact they may have on the main dataset where $\bar{V}=0$.

\subsubsection{Correlation with rf detuning.}

In analyzing EDM data we routinely search for correlations between the measured EDM and the parameters of the experiment. For the data taken with non-zero $\bar{V}$ we found such a correlation with the detuning from resonance of rf1, as measured by the $\{\nu_{\rm{rf1}}\}$ channel. In section \ref{Sec:rf phase correction} we described how an rf detuning produces an interferometer phase, and how a change in that phase due to imperfect electric field reversal results in a systematic shift to the EDM. The effect discussed here is similar but the imperfect field reversal is due to the offset voltage, $\bar{V}$, which changes the local direction of the field rather than its magnitude. The correlation observed here is with the $\{\nu_{\rm{rf1}}\}$ channel, not with the $\{\nu_{\rm{rf1}} \cdot E_0\}$ channel, which shows no such correlation.

\begin{figure}[t]
\begin{center}
\includegraphics[width=\columnwidth]{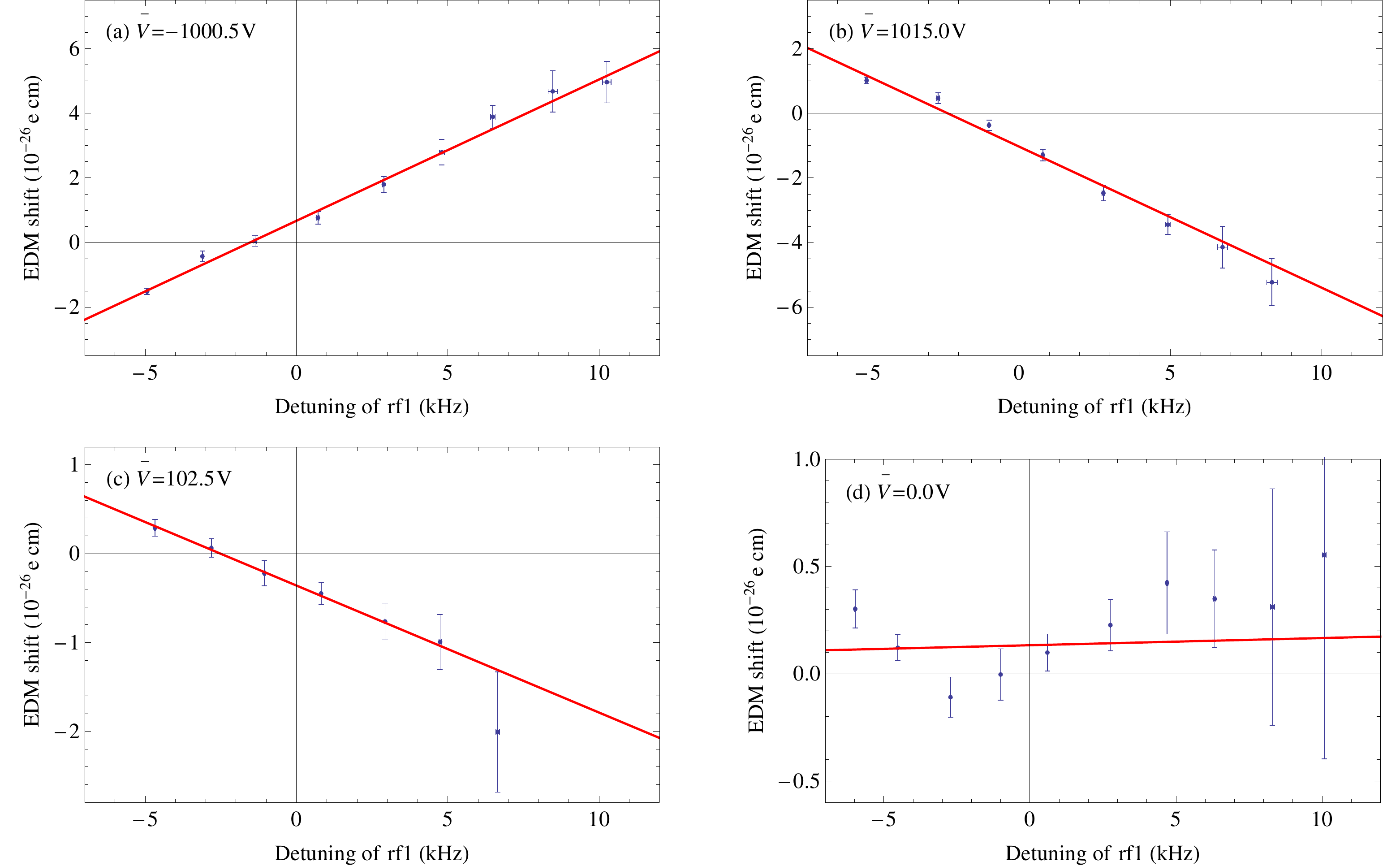}
\caption{\label{Fig:rf1Correlation} Systematic EDM shift versus detuning of rf1 for the four different values of offset voltage, $\bar{V}$. Note the change in vertical scale between the graphs. Straight line fits to the data are shown by the solid red lines. The data are consistent with this linear model in all cases. The fit gradients, in units of $10^{-28}\,e\,\rm{cm}\,\rm{kHz}^{-1}$ are (a) $43.7 \pm 1.8$, (b) $-43.7 \pm 2.0$, (c) $-14.3 \pm 2.0$, (d) $0.3 \pm 1.1$.}
\end{center}
\end{figure}

In the first rf region, the electric field varies a little across the $\sim 10$\,cm length of the molecular pulse, and so, via the Stark shift of the rf transition, the rf detuning varies across the pulse. The arrival time at the detector is almost perfectly correlated with the position of the molecules in the rf region, the ones that are ahead arriving first, so the $\{\nu_{\rm{rf1}}\}$ channel naturally varies with arrival time. By dividing up the data according to arrival time we obtain EDM values over a range of rf1 detunings. The electric field is more constant in the second rf region than the first, and we do not observe an EDM shift correlated with the $\{\nu_{\rm{rf2}}\}$ channel. Figure \ref{Fig:rf1Correlation} shows the systematic shift in the EDM with rf1 detuning for the four values of $\bar{V}$. Straight line fits to the data yield the slopes given in the caption. We see that equal and opposite values of $\bar{V}$ produce equal and opposite slopes, but the slope is not linear in $\bar{V}$. Instead, increasing $\bar{V}$ by a factor of 10, from $\approx 100$\,V to $\approx 1000$\,V, only increases the slope by a factor of 3. We find that the slopes have no dependence on the manual machine configuration.

Since the systematic shift depends on $\bar{V}$ and the gradient changes sign when $\bar{V}$ changes sign, the shift should go to zero for $\bar{V}=0$. This is indeed the case, as shown in figure \ref{Fig:rf1Correlation}(d), so there is no correction required to the main dataset. However, since there is a strong correlation with $\{\nu_{\rm{rf1}}\}$ when $\bar{V} \ne 0$, and since the scaling of the effect with $\bar{V}$ is unclear, we include a systematic uncertainty of $1.3 \times 10^{-28}\,e\,\rm{cm}$ due to a possible residual $\{\nu_{\rm{rf1}}\}$ correlation \footnote{In \cite{Hudson(1)11} we assigned a systematic uncertainty of $1.0 \times 10^{-28}\,e\,\rm{cm}$ due to a minor error in the analysis.}. This is obtained by multiplying the uncertainty in the measurement of the null gradient in figure \ref{Fig:rf1Correlation}(d), $ 1.1 \times 10^{-28}\,e\,\rm{cm}\,\rm{kHz}^{-1}$, by the measured rf1 detuning for the main dataset, 1.2\,kHz. This rf detuning differs from zero, despite the servo loop that should zero it, because there is a strong dependence of $\{\nu_{\rm{rf1}}\}$ on arrival time and because the gating of the time-of-flight profile used in the data analysis (figure \ref{fig:tof}) is different to that used by the servo loop.

\subsubsection{Remaining dependence on $\bar{V}.$}

\begin{figure}[t]
\begin{center}
\includegraphics[width=0.6\columnwidth]{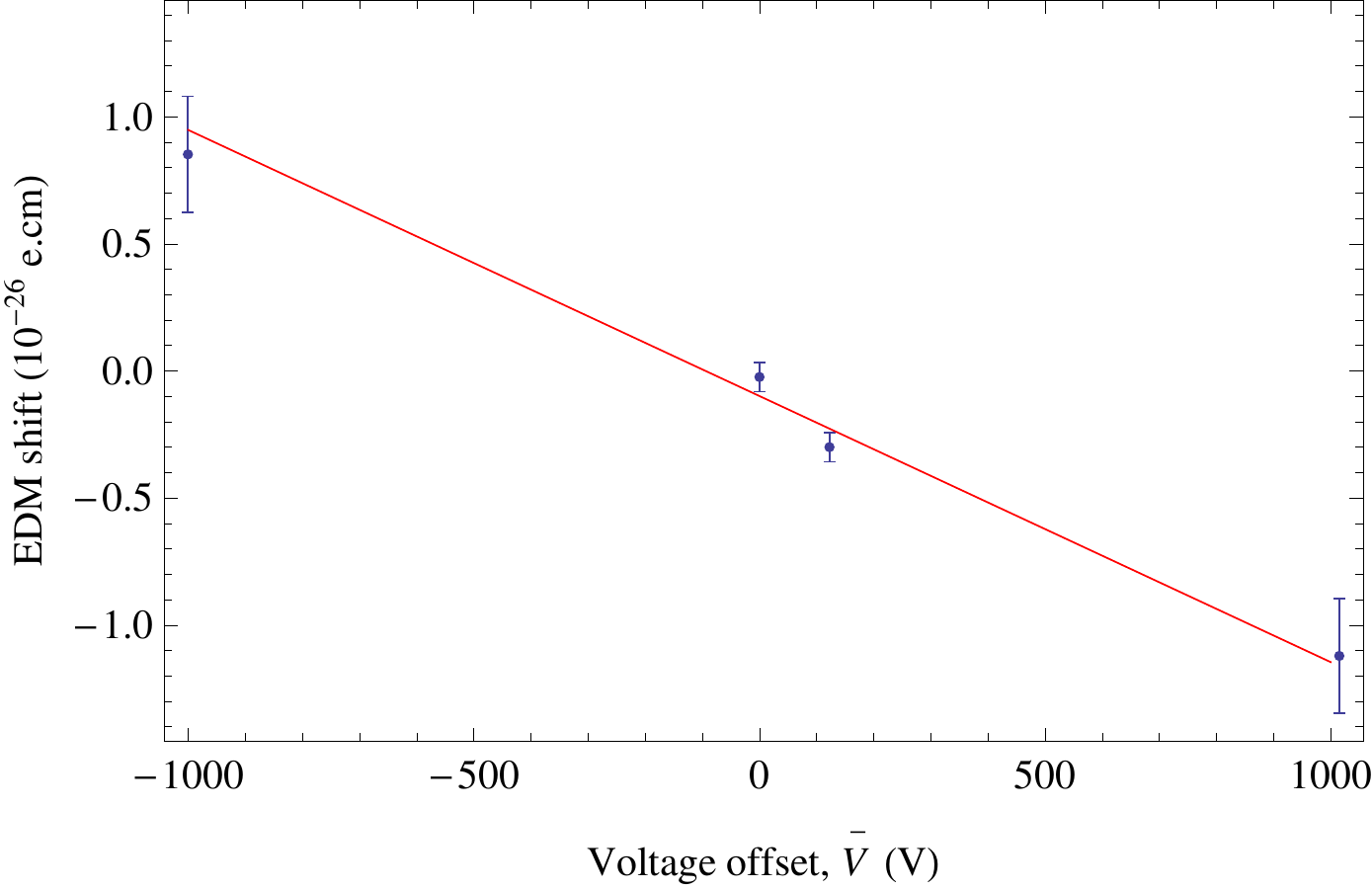}
\caption{\label{Fig:offset} EDM shift versus offset voltage.}
\end{center}
\end{figure}

It is straightforward to correct the datasets where $\bar{V} \ne 0$ for the rf1 detuning dependent shift described above. The correction is the measured gradient shown in figure \ref{Fig:rf1Correlation} multiplied by the detuning of rf1, as measured by $\{\nu_{\rm{rf1}}\}$. It is interesting to see whether there is any remaining dependence on $\bar{V}$ after making this correction. Figure \ref{Fig:offset} shows the dependence of the measured EDM on $\bar{V}$ after making this correction (in addition to the corrections explained in sections \ref{Sec:rf phase correction} and \ref{Sec:magCorrection}). We see that there is still a dependence on $\bar{V}$. The data is consistent with a linear dependence on $\bar{V}$, and a linear fit to the data gives a gradient of $(-0.099 \pm 0.016)\times 10^{-28}\,e\,\rm{cm\,V}^{-1}$. We use a high voltage probe to measure the voltage applied to each plate and thus determine the offset voltage. For the main dataset the offset voltage was measured to be smaller than 1\,V, so the resulting systematic uncertainty is $0.1 \times 10^{-28}\,e$\,cm.

\subsection{Magnetic fields}\label{Sec:magFields}

A magnetic field that changes when $E_0$ switches will produce a systematic EDM shift. For reference, a 1\,fT field along $z$ that reverses with the state of $E_0$ produces a systematic EDM shift of $4 \times 10^{-30}\,e$\,cm. The effect of a magnetic field perpendicular to $z$ is vastly smaller. This is because the Zeeman splitting between the $(F=1,M_{F}=\pm 1)$ states is exceedingly insensitive to the components of magnetic field perpendicular to the applied electric field, because of the large Stark shift, $h \Delta$, of the $(F=1,M_{F}=0)$ state relative to the other two $F=1$ states \cite{Hudson(1)02}. The effective Hamiltonian for the $F=1$ states, written in the field-free basis in order of decreasing $M_{F}$, is

\begin{equation}\label{Eq:F1Ham}
\left(\begin{array}{ccc}
h \Delta + g\mu_{B} B_{z} & g\mu_{B}B_{x}/\sqrt{2} & 0\\
g\mu_{B}B_{x}/\sqrt{2} & 0 & g\mu_{B}B_{x}/\sqrt{2} \\
0 & g\mu_{B}B_{x}/\sqrt{2}  &  h \Delta - g\mu_{B} B_{z}\\
\end{array}\right),
\end{equation}
where $B_{z}$ is the magnetic field along $z$, and $B_{x}$ is the magnetic field along $x$. To give an example, when $\Delta = 8.2$\,MHz (the Stark shift at our operating electric field), applying $B_{z}=10$\,nT produces a splitting between the $M_{F}=\pm 1$ states of 280\,Hz, but adding $B_{x}=100$\,nT to this only increases that splitting by 0.1\,mHz.

There are several ways that a magnetic field might be produced that changes when $E_0$ is switched. The first is simply that the relay and control electronics that reverse the electric field could produce a magnetic field that depends on their state. We use three magnetometers placed around the laboratory, and one between the inner and outer shield, to check for this.  One magnetometer, placed next to the relay that reverses the electric field, registers a magnetic field that changes by about 7\,nT when the relay switches. However, the relays are about 5\,m away from the machine, the field drops off rapidly with distance, and the shields have a shielding factor of a few thousand, so this field is far too small to be of concern. The other magnetometers do not register any magnetic fields that correlate with $E_0$. In particular, the average signal in the $\{E_0\}$ channel of the magnetometer situated between the shields is zero as was already discussed in section \ref{Sec:magCorrection}.

The currents that flow in the machine might also produce an $E_0$-correlated magnetic field. To reverse the electric field the power supply voltages are slowly ramped down to zero, then the plates are grounded through resistors to remove any residual charge, then the relays are switched, and finally the power supplies are slowly ramped back up again, before data taking resumes.  The whole process takes 14\,s. It is possible that the charging or discharging currents magnetize the magnetic shields, generating a magnetic field that depends on the direction of the electric field. This effect is minimized by keeping the charging current below $5\,\mu$A and by arranging for the high voltage feedthroughs to pass side-by-side through a single hole in the inner shield. As a test, we built a similar shield set-up with a similar arrangement of feedthroughs, pulsed a hundred times the normal current through these wires, and measured the magnetic field on the axis of the shield, one shield radius away from the centreline of the hole, using a fluxgate magnetometer. The change in magnetic field correlated with the reversal of the electric field was $(3.9 \pm 4.2) \times 10^{-13}$\,T. Scaling this down to the normal operating current, we deduce a systematic EDM shift due to shield magnetization of $(-0.16 \pm 0.17) \times 10^{-28}\,e\,\rm{cm}$. Since this is consistent with zero, we do not make any correction but allow a systematic uncertainty of $0.25 \times 10^{-28}\,e\,\rm{cm}$.

The leakage currents that flow during data-taking are another possible source of magnetic field. These currents are monitored continuously \cite{Sauer(1)08}, and for the main dataset the mean current that correlates with the state of $E_0$ is smaller than 1\,nA. We consider a worst-case model where a 1\,nA current flows up the edge of one plate and down the opposite edge of the opposite plate over half the length of the plates (since the high voltage feedthroughs are near the centre). The magnetic field on the axis and in the $z$-direction, averaged over the interaction region, is then 5\,fT, and generates a systematic shift of $0.2 \times 10^{-28}\,e\,\rm{cm}$. We treat this as a systematic uncertainty.

Any possible systematic shifts arising from uncontrolled magnetic fields in the directions perpendicular to $E$ must enter through an imperfect electric field reversal and so are already included in our evaluation of the systematic uncertainty. Nevertheless, as an additional check, we took a small amount of EDM data with large perpendicular magnetic fields applied. We did these tests using an applied electric field of 2.5\,kV\,cm$^{-1}$, 4 times smaller than our normal operating field. In one test we applied approximately $\pm 100$\,nT in the $y$-direction and in a second we applied approximately 500\,nT in the $x$-direction. We saw no systematic EDM shift in either case. Using these data, an estimate of $B_{x(y)}$ under normal running conditions, and a worst-case model of a phase that depends linearly on $B_{x(y)}$ but is independent of $E$ between 2.5 and 10\,kV/cm, we obtain upper limits of $0.7 \times 10^{-28}\,e$\,cm for a $B_{x}$-related systematic error in the main data, and $0.3 \times 10^{-28}\,e$\,cm for $B_{y}$. This model is very conservative, because the sensitivity to perpendicular magnetic fields is very strongly suppressed by the Stark shift, as discussed above, and this Stark shift increases from 1\,MHz at 2.5\,kV/cm to 8.2\,MHz at 10\,kV/cm.

\section{Other systematic effects}

In the previous section we discussed our empirical evaluation of the systematic uncertainties, mostly based on measurements made with various imperfections exaggerated. In this section we calculate the size of some possible effects that we could not measure directly.

\subsection{Motional magnetic field}\label{Sec:MotionalB}

The applied electric field $\vec{E}=E \hat{z}$, when transformed into the rest-frame of molecules moving with velocity $\vec{v}=v \hat{y}$, has a magnetic component $\vec{E} \times \vec{v} /c^{2} = B_{m} \hat{x}$. This motional magnetic field changes sign when $E$ is reversed, so in conjunction with a stray magnetic field, $B_{s}$, in the $x$-direction, produces a total magnetic field $B_x = B_s + B_m$ along $x$ whose magnitude changes when $E$ is reversed. This results in a systematic shift of the EDM. Fortunately, the molecule is very insensitive to fields perpendicular to $E$, as discussed in section \ref{Sec:magFields}.

The EDM shift due to the motional field is found by calculating the eigenvalues of Hamiltonian (\ref{Eq:F1Ham}), and thus determining how the splitting between the $M_{F}=\pm 1$ states changes when the electric field is reversed. In the experiment, $g=1$, $B_{z}=13.6$\,nT, $v=590$\,m\,s$^{-1}$, $E=10$\,kV\,cm$^{-1}$, and $\Delta = 8.2$\,MHz. Taking a very conservative upper limit for the stray magnetic field of $B_{s}=30$\,nT, the systematic EDM shift is only $5 \times 10^{-32}\,e\,\rm{cm}$.

\subsection{Geometric phase}\label{Sec:geometricPhase}

In addition to the dynamical phase $\phi$ given by equation (\ref{phi}), a geometric phase also contributes to the total phase of the interferometer, due to the adiabatic evolution of the molecule in fields that change their directions. Because the molecule is so insensitive to magnetic fields perpendicular to the applied electric field, as discussed in section \ref{Sec:magFields}, it is only the rotations of the electric field that need further consideration. The analysis given in reference \cite{Tarbutt(1)09} shows that this phase is equal to the solid angle swept out by the electric field vector during the period of free evolution. If the geometric phase changes when the electric field is reversed, there will be a systematic error in the measured EDM.

\begin{figure}[t]
\begin{center}
\includegraphics[width=0.6\columnwidth]{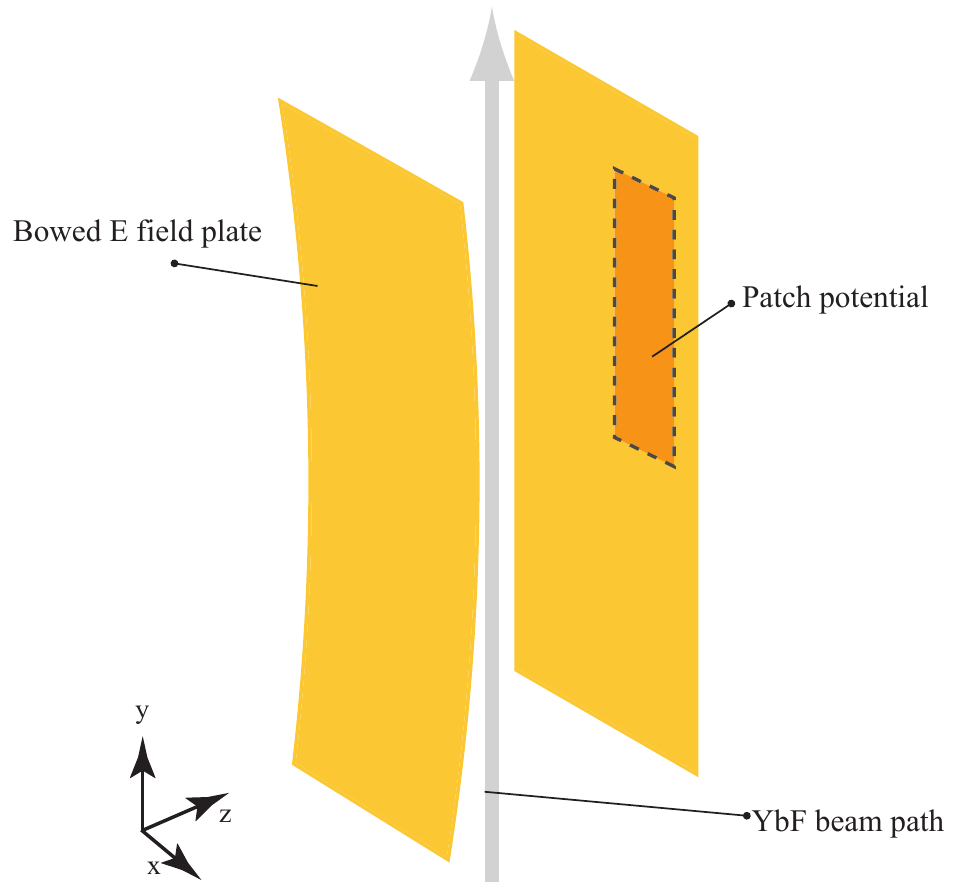}
\caption{\label{Fig:geometricPhase} Illustration of how a bowed plate and a patch potential can combine to generate a geometric phase that changes sense when the electric field is reversed.}
\end{center}
\end{figure}

There will be some rotation of the electric field direction due to the fringe fields near the edges of the plates. For the molecules that participate in the experiment this rotation is small enough to neglect. A more serious concern is the field rotation due to the bend of the plates. We have mapped the electric field magnitude along $y$ \cite{Hudson(1)07}, and this map tells us that, as a function of $y$, the field rotates about the $x$-axis by less than $\pm 0.5$\,mrad. From the geometry of the plates we expect the field rotation about the $y$-axis to be similar. This rotation does not change, either in magnitude or in sense, when the electric field reverses, and so it produces no systematic error.

Patch potentials on the plates also cause a change in the electric field direction. In the case of the patch, reversal of the electric field reverses the tilt of the electric field vector around both the $x$ and $y$ axes, so the sense of rotation remains unchanged and, once again, there is no systematic error.

The production of a systematic error requires the combination of a non-uniform gap between the plates and a patch potential. The former produces a tilt of the electric field that does not reverse with $E$, while the latter produces a tilt that does reverse with $E$, and so the combination generates a rotation that changes when $E$ is reversed. We consider a worst case example where a 1\,V patch fills the second half of the interferometer and covers half the width of the plate, as illustrated in figure \ref{Fig:geometricPhase}. In the worst case, we suppose that, as the molecules propagate between the regions where the rf pulses are applied, the electric field first rotates by 1\,mrad around $x$ because of a bend in the plates, then rotates around $y$ as the molecules enter the region of the patch, then the rotation around $x$ is reversed as the bend reverses, and finally the rotation around $y$ reverses as the molecules leave the patch area. This sweeps out a solid-angle which changes when $E$ is reversed. Averaging over the volume occupied by the molecules, this worst-case example results in a systematic error of $1 \times 10^{-30}\,e\,\rm{cm}$.\footnote{This bound is 3 times smaller than the upper bound given in \cite{Hudson(1)11} due to an improved analysis.}

\subsection{Electric field dependent sensitivity to magnetic fields}\label{Sec:eDependentBSensitivity}

At some small level, the Zeeman splitting of the $F=1$ levels must depend on the applied electric field. Considering first the parallel magnetic field, let us suppose that, at our operating electric field, the $F=1$ $g$-factor changes with electric field magnitude at the rate $\gamma$, i.e. $\gamma = (dg/d|E|)_{10\,\rm{kV/cm}}$. In combination with a change in electric field magnitude $\delta E$ on reversal, and a background magnetic field in the $z$ direction, $B_{\rm{back}}$, there will be a systematic error of size

\begin{equation}
\delta d_{\rm{e}} = \frac{\gamma \mu_{B} B_{\rm{back}} \delta E }{ E_{\rm{eff}}}.
\end{equation}
We have measured how the $g$-factor changes with applied electric field by measuring interference curves similar to the one shown in figure \ref{Fig:Interference}(a) for several values of the electric field between 3 and 14\,kV/cm. We did not find any systematic variation of the $g$-factor over this range of field, and the maximum possible variation consistent with our data is $\gamma_{\rm{max}} = 3 \times 10^{-6}$\,(V\,cm$^{-1}$)$^{-1}$. For the main dataset the $\{B_0\}$ channel gives $B_{\rm_{back}} \approx 140$\,pT. With $|\delta E| = 0.1$\,V\,cm$^{-1}$ we thus obtain a maximum systematic uncertainty from this source of $1.7 \times 10^{-31}\,e\,\rm{cm}$.

For a perpendicular magnetic field the $g$-factor is strongly suppressed by the large electric field as described in section \ref{Sec:magFields}, but for the same reason changes rapidly with $E$. The systematic shift from this effect is found by calculating how the eigenvalues of equation (\ref{Eq:F1Ham}) change for a small change in the electric field. For a perpendicular magnetic field of 30\,nT, and $|\delta E| = 0.1$\,V\,cm$^{-1}$ at $E=10$\,kV\,cm$^{-1}$ the resulting systematic EDM is only $8 \times 10^{-35}\,e\,\rm{cm}$.

\section{Additional tests}

\subsection{Non-zero channels}

As discussed in section \ref{Sec:Switches}, the analysis channels provide essential information about the experiment. The $\{E_0 \cdot B_{0}\}$ channel is hidden by the blind offset, but we study the others. In a perfect experiment only two of these other channels would be non-zero - $\{0\}$, which is simply the total signal, and $\{\delta B\}$, the change in signal due to the magnetic field step. We search the other channels looking for non-zero values. In table \ref{Tab:NonZeroChannels} we report the channels that differ from zero by more than 4 times the standard error. The channel values have been divided by the signal at the operating point on the interference curve, to give the average fractional change in the signal resulting from the corresponding switch. We note that if a channel $\{X\}$ is significantly non-zero the channel $\{X \cdot \delta B\}$ will also tend to be non-zero. We have not listed these channels since they add no extra information, except in the case of $\{\nu_{l} \cdot \delta B\}$. The manual reversals of $E$, $B$ and rf-propagation direction allow eight different ways of forming the average of a channel from the set of blocks, according to the eight ways of choosing the sign for the manual configuration. These are ${\cal E}$, ${\cal B}$, ${\cal \nu}$, ${\cal E B}$, ${\cal E \nu}$, ${\cal B \nu}$, ${\cal E B \nu}$ and unsigned. For example, we find that $\{B_{0}\}{\cal B}$ is non-zero, but $\{B_{0}\}$ is zero because the manual $B$ reversal changes its sign. We search for non-zero channels in all eight configurations. Where a channel is non-zero for more than one manual sign, table \ref{Tab:NonZeroChannels} gives the value for the most relevant sign.

\begin{table}[t]
\caption{Non-zero channels. Angle brackets indicate averaging over the arrival-time window shown in figure \ref{fig:tof}. Values give the average fractional change in the signal resulting from the corresponding switch. The bracketed number is the statistical uncertainty in the last digit. The last column gives the cause and, where appropriate, the physical value corresponding to the channel value.}\label{Tab:NonZeroChannels}
\begin{tabular}{lcl}
\br
Channel & Value ($10^{-3}$) & Cause \\
\mr
$\langle \{\nu_{l}\} \rangle$ & -11.93(10) & Other lines shift spectral peak by -6.5\,MHz from $F=0$ peak \\
$\langle \{\nu_{l} \cdot \delta B\} \rangle$ & -0.509(2) & Laser frequency detuned by -1.2\,MHz \\
$\langle \{\nu_{\rm{rf1}} \} \rangle$ & 8.87(10) & rf1 frequency detuned by 1.2\,kHz\\
$\langle \{\nu_{\rm{rf2}} \} \rangle$ & -3.22(6) & rf2 frequency detuned by -0.3\,kHz\\
$\langle \{a_{\rm{rf1}} \} \rangle$ & 1.34(6) & rf1 amplitude imperfect by 0.5\% \\
$\langle \{a_{\rm{rf2}} \} \rangle$ & 0.49(5) & rf2 amplitude imperfect by 0.2\% \\
$\langle \{ B_0 \} \rangle{\cal B}$ & 15.7(5) & Residual magnetic field of 136\,pT \\
$\langle \{E_0\} \rangle {\cal E}$ & 0.52(4) & Electric field reversal changes signal by 0.05\%\\
$\langle \{\nu_{\rm{rf1}} \cdot \nu_{\rm{rf2}}\}\rangle$ & 1.60(3) & Correlation between frequency detuning of rf1 and rf2\\
$\langle \{a_{\rm{rf2}} \cdot \Phi_{\rm{rf}} \}\rangle$ & 0.77(4) & Switching rf phase changes amplitude of rf2 by 0.3\% \\
$\langle \{\nu_{\rm{rf1}} \cdot B_{0} \}\rangle {\cal B}$ & -1.42(2) & Phase due to rf1 frequency detuning: 283\,nrad/Hz \\
$\langle \{\nu_{\rm{rf2}} \cdot B_{0} \}\rangle {\cal B}$ & 0.65(2) & Phase due to rf2 frequency detuning: -94\,nrad/Hz \\
$\langle \{\nu_{\rm{rf1}} \cdot E_0 \}\rangle {\cal E}$ & 0.16(2) & Switching $E_0$ changes $|E|$ by 0.21\,V/cm in rf1 region\\
$\langle \{\nu_{\rm{rf2}} \cdot E_0 \}\rangle {\cal E}$ & -0.17(2) & Switching $E_0$ changes $|E|$ by -0.21\,V/cm in rf2 region\\
$\langle \{\nu_{l} \cdot B_{0} \}\rangle {\cal B}$ & 0.25(2) & Not investigated\\
$\langle \{a_{\rm{rf1}} \cdot B_{0} \}\rangle {\cal B}$ & 0.14(2) & Not investigated\\
$\langle \{a_{\rm{rf2}} \cdot B_{0} \}\rangle {\cal B}$ & -0.14(2) & Not investigated \\
\br
\end{tabular}
\end{table}

The first line of the table shows that the laser frequency is significantly detuned from the peak of the spectral line. This is due to some small lines that are not resolved from the $F=0$ line. Because of these contaminant lines we do not use the $\{\nu_{l}\}$ channel to lock the laser frequency. We lock to the $F=0$ line itself, rather than the peak of the contaminated spectral line, by using the $\{\nu_{l} \cdot \delta B\}$ channel. This channel tells us how the slope of the interference curve changes when the laser frequency is stepped, which is the relevant information for the lock.

The parameters of the experiment are not set quite perfectly by the feedback loops, as shown by the next 6 lines in the table. This is no surprise since we retain all the data, including the data at the beginning of a sequence of blocks where the servo loops are still pulling the channels to zero. Also, the time-gate used in the analysis of the data (figure \ref{fig:tof}) is not the same as the gate used by the servo loops. Line 8 shows that the signal changes slightly when $E_0$ is switched. A possible reason is that the efficiency of the rf transitions changes slightly under an imperfect $E$ reversal. Note that such a change in interference amplitude, together with an uncancelled background magnetic field, contributes a $\{B_0\}\{E_0\cdot \delta B\}$ correction to the $\{E_0\cdot B_0\}$ channel in the determination of the EDM, as shown by equation (\ref{Eq:LineshapeCorrectedEDM}). As noted in section \ref{Sec:Correction} this correction is small. Line 9 of table \ref{Tab:NonZeroChannels} shows that there is a correlation between the detunings of the two rf pulses, probably because of drift in the high voltage power supplies which produces a common detuning for both rf pulses via the Stark shift of the transition frequency. Line 10 shows that the device used to introduce a $\pi$ phase change of the second rf pulse also slightly changes its amplitude. 

The channel $\{\nu_{\rm{rf1}} \cdot B_{0} \}$ is non-zero for two reasons. First, there is a contribution to this channel from the product of $\{\nu_{\rm{rf1}} \cdot \delta B\}$ and $\{B_0\}$. Once this known contribution is subtracted, the channel is still non-zero because a detuning of the rf1 frequency generates an interferometer phase. The same is true for $\{\nu_{\rm{rf2}} \cdot B_{0} \}$. Lines 13 and 14 show that the rf detunings change when $E_0$ is switched because of a small change in field magnitude which changes the Stark shift. Together, these imperfections cause the systematic error discussed in section \ref{Sec:rf phase correction}. More generally, a systematic error will result for any $X$ where $\{X \cdot B_{0} \}$ and $\{X \cdot E_0 \}$ are both non-zero. The first shows that a change in $X$ changes the phase of the interferometer, while the second shows that switching $E_0$ changes the value of $X$. We see that, although there are three more non-zero channels of the type $\{X \cdot B_{0} \}$ (last 3 lines) there are no other non-zero channels of the type $\{X \cdot E_0 \}$. Because these last three non-zero channels do not produce a systematic shift we have not investigated them further. They are probably due to the combination of non-zero $\{X \cdot \delta B\}$ and $\{B_0\}$.

In addition to the main data from the probe fluorescence detector, we also collect and analyze the data from the pump fluorescence detector, the magnetometer between the shields, the three magnetometers placed around the lab, the two leakage current monitors, and two dummy inputs (a battery and a short-circuit). Once again we search all the channels from all these data sources for signals that correlate with any of the switches (or combinations of switches). We find only a few that ideally would be zero but are not. The pump detector has a large non zero value of $\langle\{\nu_{l}\}\rangle$. This is not at all surprising -  the servo loop that locks the laser frequency uses the signal from the probe detector, and there is inevitably a difference in the Doppler shift between the two detectors because the pump and probe beams are not exactly parallel. The magnetometer between the shields registers a signal in the $\langle \{\Phi_{\rm{rf}}\} \rangle$ channel, and (marginally) in the $\langle \{a_{\rm{rf1(2)}}\} \rangle $ channels, showing that the magnetic field generated by the rf-phase switcher is different in its two states, as is the field generated by the rf amplitude-switching electronics. These fields are too small to be of concern and in any case do not depend on $E_0$. As described in section \ref{Sec:magFields}, the magnetometer that is close to the electric field relay has a non-zero $\langle\{E_0\}\rangle$ channel showing that the relays produce a magnetic field that correlates with $E_0$. However, none of the other magnetometers register this field showing that it falls off too rapidly with distance to have any significant effect on the molecules. The two leakage current monitors register small signals in their $\langle \{B_{0}\}\rangle$ channels, which, though baffling, are much too small to be of concern. All channels of the dummy inputs (other than $\langle \{0\} \rangle$) are zero showing that there is no bias in the data collection or analysis procedures.

\subsection{Other correlations}

The polarization angles of the linearly-polarized pump and probe laser beams are changed randomly from one block to the next, and their values for each block are recorded. We do not find any dependence on the two polarization angles when the measured EDMs are divided up according to these angles.

We also randomly change the phase difference between the two rf pulses from one block to the next, as described in section \ref{Sec:lineshape}. Again, we see no dependence of the EDM on the value of this phase difference.

Finally, we do the analysis separately for each of the 8 manual machine configurations. The EDMs obtained are all consistent with one another.

\section{Result}\label{Sec:Result}

\begin{table}[tb]
\caption{Statistical uncertainty, contributions to the systematic uncertainty, and the total systematic uncertainty.}\label{Tab:Uncertainties}
\begin{tabular}{ll}
\br
Source & Uncertainty ($10^{-28} e\,\rm{cm}$)\\
\mr
Total statistical uncertainty & 5.7 \\
\hline
Uncorrected effects due to electric field asymmetry, $\delta E$ & 1.1\\
Residual correlation with rf1 detuning & 1.3\\
Uncertainty due to residual voltage offset, $\bar{V}$ & 0.1\\
Leakage currents & 0.2\\
Shield magnetization & 0.25\\
Geometric phase & 0.01\\
Motional magnetic field & 0.0005\\
\hline
Total systematic uncertainty & 1.7\\
\br
\end{tabular}
\end{table}

As described in section \ref{EDM mean and error}, we find the central value of the EDM using the 5\% trimmed mean of the set of EDMs measured by each block, each corrected for the rf detuning correction (section \ref{Sec:rf phase correction}) and the magnetic field correction (section \ref{Sec:magCorrection}). The statistical uncertainty is found from this set of corrected blocks using the bootstrap method. Table \ref{Tab:Uncertainties} gives this statistical uncertainty and summarizes the contributions to the systematic uncertainty. The total systematic uncertainty is calculated by adding these contributions in quadrature. The final result is

\begin{equation}
d_{\rm{e}} = (-2.4 \pm 5.7_{\rm{stat}} \pm 1.7_{\rm{syst}}) \times 10^{-28} e\,\rm{cm}. \nonumber
\end{equation}

To calculate confidence intervals on the value of $|d_{\rm{e}}|$, we create the distribution of $|d_{\rm{e}}|$ values using the bootstrap method, and integrate this distribution from zero to $d_{\rm{c,stat}}$ such that the integral is $c$\%. This $d_{\rm{c,stat}}$ is the statistical bound of the $c$\% confidence interval. The systematic uncertainty for this confidence level $d_{\rm{c,syst}}$ is derived from a Gaussian distribution with zero mean and a standard deviation of $1.7 \times 10^{-28} e\,\rm{cm}$. The upper bound on $|d_{\rm{e}}|$ at this confidence level is then taken as $\sqrt{d_{\rm{c,stat}}^{2} + d_{\rm{c,syst}}^{2}}$. Table \ref{Tab:UpperBounds} gives these upper bounds for various confidence levels.

As mentioned in section \ref{Sec:Overview}, we have interpreted our measurement in terms of the electron EDM alone. There can also be contributions to the EDM of the YbF molecule from P,T-violating electron-nucleon interactions. To disentangle these various possible contributions requires measurements to be made in different systems where the relative sensitivities to these contributions are different. To obtain the EDM of the YbF molecule at 10\,kV/cm from our electron EDM value, multiply by $E_{\rm{eff}}/E = -1.45 \times 10^{6}$.

\section{Conclusions and outlook}

\begin{table}[!t!b]
\caption{Upper bounds for $|d_{\rm{e}}|$ at various confidence levels.}\label{Tab:UpperBounds}
\begin{indented}
\item[]\begin{tabular}{ll}
\br
Confidence level (\%) & Bound ($10^{-28} e\,\rm{cm}$)\\
\mr
68.3 & 6.5\\
90.0 & 10.6\\
95.8 & 13.1\\
99.5 & 18.5\\
\br
\end{tabular}
\end{indented}
\end{table}

In this paper, we have presented a detailed account of our measurement of the electron EDM using YbF molecules, focussing on the data analysis and the evaluation of the uncertainties.  At present, our limiting uncertainty is statistical. This is the first time that the precision of a molecular measurement of the electron EDM has exceeded that of the best atomic measurement. We anticipate a series of new measurements, of increasing precision, using this new method. By separating the rf transmission line from the electric field plates, we will use the length of the machine more efficiently, and by shortening the rf pulses we will be able to use a higher fraction of the available molecules. With these upgrades we expect to reduce the statistical uncertainty by a factor of 3. After this, we aim to reduce the statistical uncertainty by a further factor of 10 or more with the use of a cryogenic buffer gas source of YbF \cite{Skoff(1)11}, where the flux will be 10 times higher and the speed 3 times lower \cite{Barry(1)11} than for the present source. At this higher sensitivity, it will be necessary to reduce the magnetic noise, which can be done by adding a third layer of magnetic shielding. A set of spin-exchange relaxation-free alkali vapour magnetometers can be placed inside the machine to greatly improve the magnetometry. In the present measurement we corrected for a systematic shift due to the rf-detuning induced phases. The size of the correction was approximately equal to the statistical uncertainty. We expect to reduce this systematic effect by at least a factor of 100 by shortening the rf pulses by a factor of 10, improving the control over the rf polarization, and improving the electric field reversal by a factor of 10. The latter can be done through Ramsey interferometry of the hyperfine interval to compare the Stark shift in the two electric field states, as we demonstrated in \cite{Tarbutt(1)08}. This improved control over the electric field reversal will also reduce the leading systematic uncertainties to below $10^{-29}\,e$\,cm. In the longer term, a further large improvement in statistical sensitivity seems possible through a combination of a thermal cryogenic source of very slow molecules \cite{Lu(1)11} combined with direct laser cooling \cite{Shuman(1)10}. The radiative properties of the A--X transition of YbF make it a suitable molecule for laser cooling \cite{Zhuang(1)11}. If these developments can be implemented, a measurement at the $10^{-30}\,e$\,cm sensitivity level is within reach.

Several new atomic and molecular EDM experiments are now underway or are being developed, as recently reviewed in \cite{Commins(1)10}, and they too are expected to reach similar levels of precision. Together, these new experiments will probe deep into the region where a non-zero EDM should be found, if current theories that extend the standard model are correct \cite{Pospelov(1)05}.

\ack
We acknowledge the contributions of P. Condylis and H. Ashworth. We are grateful for engineering support from J. Dyne and V. Gerulis. This
work was supported by the UK research councils STFC and EPSRC, and by the Royal Society.

\newpage

\appendix
\section{Determining the EDM in an imperfect experiment}\label{AppA}

In section \ref{Sec:Correction} we derived an expression to extract the phase shift associated with the electric-field switch from the measured analysis channels. We saw that if one allows for imperfections in the experiment -- namely that the $E_0$ and $B_0$ switches change the amplitude of the signal and there is an uncancelled background field $B_{\rm{back}}$ -- then terms additional to $\{E_0 \cdot B_0\}$ are needed to extract the EDM phase shift. Here we consider more general imperfections, and derive the corresponding correction terms. Following the treatment of section  \ref{Sec:Correction} we define our signal as
\begin{equation}
S(\widehat{E_0}, \widehat{B_0}, \widehat{\delta B}, \widehat{Q}) = A f(\Theta)\,
\end{equation}
where we have introduced an extra switched parameter with state $\widehat{Q}$. This can be any of the other parameters that are switched in the experiment. If both the amplitude and phase of the signal depend on this switch then a correction will be required. We suppose a signal amplitude of the form
\begin{equation}
\fl A = 1 +  \delta_{b} \widehat{B_0} + \delta_{e} \widehat{E_0}  + \delta_{q} \widehat{Q} + \delta_{b,e} \widehat{B_0} \widehat{E_0} + \delta_{b,q} \widehat{B_0} \widehat{Q} + \delta_{e,q} \widehat{E_0} \widehat{Q} + \delta_{b,e,q} \widehat{B_0} \widehat{E_0} \widehat{Q}\,
\end{equation}
meaning that the amplitude can depend on any combination of the switched parameters, apart from the small calibration step $\widehat{\delta B}$. We have renamed the imperfection parameters, as compared to section \ref{Sec:Correction}, to make the notation more straightforward. We define the phase function
\begin{equation}
f(\Theta)=(\phi_{\rm{back}} \widehat{B_0}+\phi_{\delta B} \widehat{\delta B}+\phi_{EDM} \widehat{B_0} \widehat{E_0}+\phi_Q \widehat{B_0} \widehat{Q}) - \beta.
\end{equation}
This phase function is essentially the same as that used in section \ref{Sec:Correction} except for the addition of a phase, $\phi_{Q}$, which depends on the switch state $\widehat{Q}$. To simplify the derivation, we have defined $f(\Theta)$ to be linear in the switched parameters directly, rather than carrying through an arbitrary function and linearizing as in section \ref{Sec:Correction}. The parameter $\beta$ is proportional to $f(\phi_{B_0})$ of section \ref{Sec:Correction}.

We wish to determine the ratio of $\phi_{EDM}$ to $\phi_{\delta_B}$. It is straightforward, though quite tedious, to show that
\begin{equation}
\frac{\phi_{EDM}}{\phi_{\delta_B}} =  \frac{1}{N} (T_1 + T_2 + T_3 + \beta(T_4 + T_5)) \ ,
\end{equation}
where
\begin{eqnarray}
\fl N =\!\{\delta B\}^3\!+\!2 \{\delta B \cdot Q\} \{\delta B \cdot E_0\} \{\delta B \cdot E_0 \cdot Q\}\!-\!\{\delta B\} \left( \{\delta B \cdot Q\}^2\!+\!\{\delta B \cdot E_0\}^2\!+\!\{\delta B \cdot E_0 \cdot Q\}^2\right) \nonumber\\
\fl T_1 = \left(\{\delta B \}^2 - \{\delta B \cdot Q\}^2\right) \{B_0 \cdot E_0\} \nonumber\\
\fl T_2 = \left(\{\delta B \cdot Q\} \{B_0 \cdot Q\} - \{B_0\} \{\delta B\}\right) \{\delta B \cdot E_0\} \nonumber\\
\fl T_3 = \left(\{B_0\} \{\delta B \cdot Q\} - \{\delta B\} \{B_0 \cdot Q\}\right)\{\delta B \cdot E_0 \cdot Q\} \nonumber\\
\fl T_4 = \{\delta B\} \{\delta B \cdot B_0 \cdot E_0\} - \{\delta B \cdot B_0\} \{\delta B \cdot E_0\} - \{\delta B \cdot B_0 \cdot Q\} \{\delta B \cdot E_0 \cdot Q\} \nonumber\\
\fl T_5 = \frac{ \{\delta B \cdot Q\} }{ \{\delta B\} } \left( \{\delta B \cdot E_0\} \{\delta B \cdot B_0 \cdot Q\} - \{\delta B \cdot Q\} \{\delta B \cdot B_0 \cdot E_0\} + \{\delta B \cdot B_0\} \{\delta B \cdot E_0 \cdot Q\}\right)\nonumber
\end{eqnarray}

None of the additional correction terms were used in the analysis presented in this paper, as they were all negligibly small.

\end{document}